\documentclass[a4paper,11pt]{article}

\usepackage[a4paper,left=2.73cm,right=2.7cm,top=3cm,bottom=3.5cm]{geometry}

\usepackage{amsmath,amssymb}

\usepackage[colorlinks=true,linktocpage=true,linkcolor=blue,citecolor=blue,urlcolor=blue]{hyperref}

\usepackage[sort&compress,merge,numbers]{natbib}

\addtocontents{toc}{\protect\setcounter{tocdepth}{2}}
\numberwithin{equation}{section}

\usepackage{epsfig}
\usepackage{graphicx}
\usepackage{epstopdf}
\usepackage{caption}
\usepackage{subcaption}

\def\ds{\,\delta_\sigma}

\def\ds{{\textbf{ds}^2}}
\def\k{{\textbf{k}}}
\def\m{{\textbf{m}}}

\def\a{{\textbf{a}}}
\def\b{{\textbf{b}}}

\newcommand{\ku}[1]{\textbf{k}^#1\,}

\newcommand{\kpu}[1]{{\textbf{k}_\perp}^#1\,}

\newcommand{\be}{\begin{equation}}
\newcommand{\ee}{\end{equation}}
\newcommand{\beq}{\begin{equation}}
\newcommand{\eeq}{\end{equation}}
\newcommand{\bal}{\begin{align}}
\newcommand{\eal}{\end{align}}
\newcommand{\bse}{\begin{subequations}}
\newcommand{\ese}{\end{subequations}}

\begin{document}

\begin{titlepage}

\thispagestyle{empty}

\begin{flushright}
\hfill{}
\end{flushright}

\vspace{40pt}  
	 
\begin{center}

{\huge \textbf{Extremal Black Hole Horizons}}
	\vspace{30pt}
		
{\large \bf Jay Armas$^{1}$, Troels Harmark$^{2}$ and Niels A. Obers$^{2}$}
		
\vspace{25pt}

{$^{1}$Universit\'e Libre de Bruxelles (ULB) and International Solvay Institutes,\\
Service  de Physique Th\'eorique et Math\'ematique, \\
Campus de la Plaine, CP 231, B-1050, Brussels, Belgium.}
\vspace{15pt}

{$^{2}$The Niels Bohr Institute, University of Copenhagen\\
Blegdamsvej 17, 2100 Copenhagen \O, Denmark.}\\

\vspace{20pt}
{\tt 
\href{mailto:jarmas@ulb.ac.be}{jarmas@ulb.ac.be}, 
\href{mailto:jtarrio@ulb.ac.be}{harmark@nbi.dk},
\href{mailto:jtarrio@ulb.ac.be}{obers@nbi.dk}
}

\vspace{40pt}
				
\abstract{
Using the blackfold effective theory applied to extremal Kerr branes we provide evidence for the existence of new stationary extremal black hole solutions in asymptotically flat spacetime with both single and multiple disconnected horizons. These include extremal doubly-spinning black rings, black saturns, di-rings and bi-rings in five spacetime dimensions as well as extremal Myers-Perry black holes and black saturns in dimensions greater than five. Some of these constructions constitute the first examples of black hole solutions with extremal disconnected horizons in vacuum Einstein gravity.

}

\end{center}

\end{titlepage}

\tableofcontents

\hrulefill
\vspace{10pt}

\section{Introduction} \label{sec:intro}
Despite enormous progress \cite{Myers:1986un, Emparan:2001wn, Emparan:2003sy, Pomeransky:2006bd, Emparan:2007wm, Elvang:2007rd, Evslin:2007fv, Iguchi:2007is, Elvang:2007hs, Izumi:2007qx, Figueras:2008qh, Emparan:2009cs, Emparan:2009at, Emparan:2009vd, Kleihaus:2012xh, Dias:2014cia, Kleihaus:2014pha, Emparan:2014pra, Figueras:2015hkb, Armas:2015kra, Armas:2015nea, Figueras:2017zwa}, the solution space of asymptotically flat black holes and its properties in vacuum Einstein gravity in $D\ge5$ is still far from being completely understood.\footnote{There is also is a very rich phase structure for black holes in vacuum Einstein gravity with Kaluza-Klein asymptotics (see Ref.~\cite{Harmark:2007md, Horowitz:2012nnc}).} This is in particular the case as $D$ increases or when one considers multi-spinning black holes and their extremal limits. 

The recent development of numerical techniques has given a better grasp of the phase structure of singly-spinning black holes \cite{Kleihaus:2012xh, Dias:2014cia, Kleihaus:2014pha, Emparan:2014pra, Figueras:2015hkb, Figueras:2017zwa} but even in this simpler situation most studies are limited to a few number of topologies and to the assumption of single horizons. 

Exact and analytic solutions are also scarce. The Myers-Perry solution is the only known vacuum black hole solution in $D\ge6$ \cite{Myers:1986un}. In $D=5$, following the discovery of the singly-spinning black ring \cite{Emparan:2001wn} and the realisation that the rod structure could be used to generate new solutions \cite{Harmark:2004rm, Harmark:2005vn}, many new regular black hole spacetimes were found, with and without disconnected horizons \cite{Pomeransky:2006bd,Elvang:2007rd, Evslin:2007fv, Iguchi:2007is, Elvang:2007hs, Izumi:2007qx}. This, however, is far from exhausting all possible geometries and topologies in higher dimensions.

Hawking's topology theorem states that horizon cross sections in $D=4$ must have $\mathbb{S}^{2}$ topology \cite{Hawking:1973uf}. The generalisation of this theorem to higher dimensions states that cross sections of stationary horizons are of positive Yamabe type \cite{Galloway:2005mf}. However, this only implies loose restrictions on horizon topology, as nor have all topologies satisfying this condition been classified neither is it known that for every topology satisfying this condition there will be a corresponding regular black hole solution.

The difficulty in understanding gravity in higher dimensions has prompted the development of perturbative but analytic methods. The blackfold effective theory \cite{Emparan:2009cs, Emparan:2009at} has proven to be a powerful tool to scan for the existence of new black hole solutions in $D\ge5$. In particular, it has shown to be a framework that realises many new horizon topologies, whose cross section is of positive Yamabe type \cite{Emparan:2009vd, Galloway:2011np, Armas:2015kra, Armas:2015nea}. It has also given several examples of new classes of black hole solutions, which despite sharing the same topology, differ in their geometry and asymptotic charges. Such is the case for black rings, helical black rings and helicoidal black rings in $D=6$ which all have topology $\mathbb{S}^{1}\times\mathbb{S}^{3}$ but not the same properties \cite{Emparan:2009vd, Armas:2015nea}.

Except for the Myers-Perry \cite{Myers:1986un} and the doubly-spinning black ring solution \cite{Pomeransky:2006bd}, each black hole horizon of all the known exact solutions in $D=5$ mentioned above only has one non-vanishing angular momentum. In-depth studies of the black saturn solution \cite{Elvang:2007rd, Eckstein:2013fra}, concentric black ring solution (di-rings) \cite{Evslin:2007fv, Iguchi:2007is} and orthogonal black ring solution (bi-rings) \cite{Elvang:2007hs, Izumi:2007qx} have shown that they do not exhibit regular extremal limits, or extremal limits that do not involve naked singularities. The only known solutions that admit regular extremal limits are the Myers-Perry and the doubly-spinning black ring solution. Proceeding by induction, this supports the statement that two non-vanishing and independent angular momenta are required in order to have an extremal stationary asymptotic flat black hole solution in $D=5$.\footnote{This claim does not contradict earlier literature since there is no complete classification of extremal vacuum asymptotically flat black holes in $D\ge5$.}

Following this line of thought, this paper uses the blackfold effective theory for stationary Kerr-branes developed in \cite{Armas:2011uf, Armas:2013hsa, Armas:2014rva, toappear} in order to provide strong evidence for the existence of many doubly- and multi-spinning extremal black holes in $D\ge5$. These black holes are ultraspinning in one or more angular directions but have finite angular momentum in one direction. This finite angular momentum satisfies the Kerr bound, allowing these black holes to have regular extremal limits. The latter thus implies that the effective blackfold worldvolume has co-dimension 3. In particular, besides showing that the specific cases of extremal doubly-spinning black rings and Myers-Perry black holes can be reproduced, it is shown that by allowing each horizon to be doubly-spinning, black saturns, di-rings and bi-rings in $D=5$ can have regular extremal limits. This is done by perturbatively placing a black ring horizon around an extremal doubly-spinning Myers-Perry black hole or black ring solution. Some of these constructions are generalised to $D\ge6$.

This paper is one in a series of two, where new doubly- and multi-spinning black hole solutions with disconnected horizons are found using the blackfold effective theory for Kerr branes. The work presented here focuses on extremal black holes, while \cite{toappear} presents a detailed study of finite temperature black holes. This paper is organised as follows. In Sec.~\ref{sec:kerrbrane}, the effective blackfold theory for stationary Kerr branes is introduced along with the notation required for the later sections of this work. In Sec.~\ref{sec:bh5d}, we focus on extremal black holes in $D=5$ and study their phase diagram. In Sec.~\ref{sec:bhd} we study a few examples of extremal black holes in $D\ge6$, in particular, we provide a brief study of higher dimensional black saturns. In Sec.~\ref{sec:conclusions} we conclude with open questions and future research directions.

\section{Effective theory for extremal stationary Kerr branes} \label{sec:kerrbrane}
In this section we introduce the effective theory for stationary Kerr branes as developed in \cite{toappear}, which is heavily based on the results of \cite{Armas:2011uf, Armas:2013hsa, Armas:2014rva}. We introduce the finite temperature case, valid for any codimension, to begin with and obtain the extremal version by taking the zero temperature limit, showing that only the codimension 3 case is regular in the limit. This effective theory consists of bending Kerr branes (Kerr black holes with additional $p$ flat spatial directions) over an arbitrary submanifold. This theory allows to construct new extremal solutions for which one of its angular momenta is finite but the remaining ones are ultraspinning. In order to consider further finite angular momenta, one must deal with the effective theory of Myers-Perry branes, which is considered in \cite{toappear}.


\subsection{Geometry and notation} 
We consider a $(p+1)$-dimensional submanifold $\mathcal{W}_{p+1}$ with Lorentzian signature immersed in a $D$-dimensional background spacetime endowed with metric $g_{\mu\nu}(x^{\mu})$ where $x^{\mu}$ are background coordinates. Greek indices $\mu,\nu,...$ label spacetime indices. The submanifold has co-dimension $(n+2)$ so that the spacetime dimension can be written as $D=n+p+3$. The location of the surface is described by the embedding map $X^{\mu}(\sigma^{a})$ where $\sigma^{a}~,~a=0,...,p$ are coordinates on the submanifold. Given the embedding map, one may explicitly define a set of tangent vectors ${e^{\mu}}_a$ and implicitly define a set of normal vectors ${n^{\mu}}_i$ according to
\beq
{e^{\mu}}_a=\partial_a X^{\mu}~~,~~g_{\mu\nu}{e^{\mu}}_a{n^{\nu}}_i=0~~,~~g_{\mu\nu}{n^{\mu}}_j{n^{\nu}}_i=\delta_{ij}~~,
\eeq
where the indices $i,j,..$ label the normal $(n+2)$ directions. With this we may define the induced metric $\gamma_{ab}$, extrinsic curvature ${K_{ab}}^{i}$ and the spin connection ${\omega_{a}}^{ij}$ as
\beq
\gamma_{ab}=g_{\mu\nu}(X){e^{\mu}}_a{e^{\nu}}_b~~,~~{K_{ab}}^{i}={n_\mu}^{i}\nabla_a {e^{\mu}}_b~~,~~{\omega_{a}}^{ij}={n_\mu}^{i}\nabla_a n^{\mu j}~~,
\eeq
where $\nabla_a$ is the Christofell connection compatible with both $g_{\mu\nu}$ and $\gamma_{ab}$ and does not act on the normal indices $i,j$. Following \cite{Armas:2014rva}, we focus on background geometries with a $q$-number of $U(1)$ isometries corresponding to a $q$-number of transverse spin planes. We define the Levi-Civita symbol on each spin plane by $\epsilon^{ij}_{(q)}$ so that the spin connection can be decomposed according to
\beq
{\omega_{a}}^{ij}=\sum_{q}\omega_{a}^{(q)}\epsilon^{ij}_{(q)}~~,~~\omega_{a}^{(q)}=\frac{1}{2}\epsilon_{ij}^{(q)}{\omega_{a}}^{ij}~~,
\eeq
where $\omega_{a}^{(q)}$ is the normal fundamental one-form on a given spin plane. This is enough to construct a free energy functional that describes the stationary sector of the effective theory governing long-wavelength deformations of the Kerr brane.


\subsection{Effective free energy functional}
In order to construct an effective free energy functional for deformations of Kerr branes, we assume the existence of a perturbative parameter $\varepsilon\ll1$ which, for a surface that is bent, is typically associated with the brane thickness $r_0$ and the extrinsic curvature scale $R$ of the submanifold such that $\varepsilon\equiv r_0/R$. The precise form of $\varepsilon$ is case dependent \cite{Armas:2015kra} but will always involve the temperature $T$ or the transverse angular velocity $\widehat\Omega$ of the configuration and some other scale related to the deformation that is being applied to the brane. Keeping this in mind, the effective free energy can be constructed in a derivative expansion and to first order in $\varepsilon$ it takes the form \cite{Armas:2013hsa,Armas:2014rva}
\beq \label{eq:eff}
\mathcal{F}[T,\Omega^{(l)},\hat \Omega, X^{\mu},g_{\mu\nu}]=-\int_{\mathcal{B}_p}\sqrt{|\gamma|}\left(P(\mathcal{T},\omega)+2\mathcal{J}(\mathcal{T},\omega)u^{a}\omega_{a}+...\right)~~,
\eeq
where the \emph{dots} represent higher-order corrections which we will not explicitly deal with in this paper. All quantities appearing in \eqref{eq:eff} are evaluated on the hypersurface $x^\mu=X^{\mu}$. In \eqref{eq:eff} we have introduced $\mathcal{B}_p$, which represents the spatial part of the submanifold worldvolume after Wick rotating the induced metric $\gamma_{ab}$ and integrating over the time coordinate with period $2\pi/T$. The quantities $\mathcal{T}$ and $\omega$ denote the local brane temperature and local transverse angular velocity and are related to the global (constant) temperature $T$ and normal angular velocity $\widehat\Omega$ according to \cite{Armas:2014rva}
\beq
\mathcal{T}=\frac{T}{\textbf{k}}~~,~~\omega=\frac{\widehat\Omega}{\textbf{k}}~~,
\eeq
where $\textbf{k}$ is the modulus of the worldvolume Killing vector field $\textbf{k}^{a}$ which we take to be of the general form
\beq
\textbf{k}^{a}\partial_a=\partial_\tau+\sum_{l}\Omega^{(l)}\partial_{\phi_{(l)}}~~,
\eeq
with $\tau$ being the time coordinate on the submanifold and $\Omega^{(l)}$ the angular velocity on each angular coordinate $\phi_{(l)}$ of the worldvolume. We assume that this Killing vector field can be pushed-forward to a background Killing vector field, i.e., $\textbf{k}^{\mu}=\textbf{k}^{a}{e^\mu}_a$. We also assume that the transverse angular velocity $\widehat\Omega$ is the angular velocity associated with a background transverse Killing vector field $\textbf{k}_\perp^{\mu}\partial_\mu=\widehat\Omega \partial_{\psi}$ where $\psi$ labels the transverse angular coordinate. 

The stress tensor that follows from \eqref{eq:eff} by varying with respect to $\gamma_{ab}$ is that of an effective fluid described by a local temperature $\mathcal{T}$ and chemical potential $\omega$. The vector $u^{a}$ appearing in \eqref{eq:eff} is the velocity $u^{a}=\textbf{k}^{a}/\textbf{k}$ of the fluid living on the submanifold, unit normalised such that $u^{a}u_a=-1$. Furthermore, we assume that the spin-orbit coupling behaves as $\mathcal{J}(\mathcal{T},\omega)u^{a}\omega_{a}\sim\mathcal{O}\left(\varepsilon\right)$. Note that since the Kerr brane is singly-spinning it can only have transverse angular momentum in one spin plane, therefore we have omitted the index $(q)$ from $\omega_{a}^{(q)}$. The scalars $P$ and $\mathcal{J}$ appearing in \eqref{eq:eff} denote the pressure density and the density of transverse angular momentum, respectively, of the fluid and are given by \cite{Armas:2011uf}
\beq \label{eq:PJ}
P=-\frac{\Omega_{(n+1)}}{16\pi G}r_0^{n}\left(1+\frac{\widehat b^2}{r_0^2}\right)~~,~~\mathcal{J}=\frac{\Omega_{(n+1)}}{8\pi G}r_0^{n}~\widehat b \left(1+\frac{\widehat b^2}{r_0^2}\right)~~,
\eeq
where $\Omega_{(n+1)}$ is the volume of the transverse $(n+1)$-sphere, $G$ is Newton's constant, $r_0$ is the brane thickness and $\widehat b$ the density of intrinsic rotation. The pressure and transverse angular momentum densities satisfy the first law of termodynamics $dP=-sd\mathcal{T}+\mathcal{J}d\omega$. In the stationary case these are related to the global potentials $T,\widehat\Omega$ according to \cite{toappear}
\beq \label{eq:solrb}
\frac{r_0}{\textbf{k}}=\frac{2(n-1)\pi T\!\!+\!\sqrt{4\pi^2T^2\!-\!\!(n-2)n \widehat\Omega^2}}{2(4\pi^2T^2+\widehat\Omega^2)}~,~\frac{\widehat b}{\textbf{k}}=\frac{n\widehat\Omega^2\!+\!2\pi T\left(2\pi T\!-\!\sqrt{4\pi^2T^2\!-\!(n-2)n \widehat\Omega^2}\right)}{2\widehat\Omega(4\pi^2T^2+\widehat\Omega^2)}~,
\eeq
with the condition that $4\pi^2T^2\!-\!\!(n-2)n \widehat\Omega^2\ge0$ which is a remnant of the Kerr bound of the Kerr brane. In the limit in which the Kerr brane is ultraspinning, i.e. $\widehat b/r_0\gg1$, these quantities agree with those obtained in \cite{Armas:2015nea} by integrating out cross-sections of the worldvolume of non-rotating branes. Also, when sending $\widehat\Omega\to0$, \eqref{eq:solrb} agrees with the corresponding quantities for non-spinning branes \cite{Emparan:2009at}.


\subsection{The extremal limit}
Taking the extremal limit amounts to sending $T\to0$. Setting $T=0$ in \eqref{eq:solrb} leads to 
\beq \label{eq:solrb0}
r_0|_{T=0}=\textbf{k}\frac{\sqrt{(2-n)n}}{2\widehat\Omega}~~,~~\widehat b|_{T=0}=\textbf{k}\frac{n}{2\widehat\Omega}~~,
\eeq
where, without loss of generality, we have assumed that $\widehat \Omega>0$. We see from \eqref{eq:solrb0} that the reality of $r_0$ and the requirement of a non-vanishing thickness\footnote{A vanishing thickness implies a vanishing mass and here we wish to construct black holes with non-vanishing mass. In fact, for $n=2$, this limit corresponds to a naked singularity of the Kerr string.} implies that $n=1$. Therefore, extremal Kerr branes can only be used to construct new solutions with $n=1$. We assume this to be the case in the remaining of this paper. Introducing \eqref{eq:solrb0} into the effective free energy \eqref{eq:eff} yields
\beq \label{eq:eff0}
\mathcal{F}[\Omega,\hat \Omega, X^{\mu},g_{\mu\nu}]|_{T=0}=\frac{1}{4\widehat\Omega G}\int_{\mathcal{B}_p}\sqrt{|\gamma|}\textbf{k}\left(1-\frac{2}{\widehat \Omega}\textbf{k}u^{a}\omega_a\right)~~,
\eeq
where we have used that $\Omega_{(2)}=4\pi$ since we have set $n=1$ in \eqref{eq:PJ}. From \eqref{eq:eff0} we observe that the effective free energy is proportional to $\textbf{k}$, which is a result that holds even if a non-zero temperature had been considered. This implies that the zeroth order equilibrium condition that is obtained by solving the equations of motion that arise from varying \eqref{eq:eff0} is the same as that for finite temperature uncharged and non-spinning branes \cite{Emparan:2009at}. Differences appear when corrections are considered. On the other hand, the thermodynamic properties of the solutions are completely different. Configurations found in previous works \cite{Emparan:2009vd, Armas:2015kra, Armas:2015nea} may also exist at extremality as long as there are no symmetry constraints.\footnote{In particular, Ref.~\cite{Emparan:2009vd} shows the possibility of finding non-extremal helical black rings in $D=5$ but also shows the impossibility of having extremal helical black rings as they break the necessary $U(1)$ symmetry required for a regular extremal limit.} 

\subsection{Equations of motion}
The equations of motion that arise from \eqref{eq:eff0} can be obtained by either performing a diffeomorphism of the background coordinates such that $\delta g_{\mu\nu}=2\nabla_{(\mu}\xi_{\nu)}$ for some infinitesimal vector field $\xi^{\mu}$ keeping the embedding map fixed \cite{Armas:2017pvj} or by slightly deforming the embedding map $X^{\mu}\to X^{\mu}+\delta X^{\mu}$ and requiring invariance of the free energy \eqref{eq:eff0} under infinitesimal rotations of the normal vectors \cite{Armas:2013hsa, Armas:2014rva}. In both cases, the global potentials $\Omega_{(l)},\widehat\Omega$ are kept fixed under the variation. The non-trivial equation of motion and boundary conditions for normal deformations read \cite{Armas:2013hsa}\footnote{The parallel projection of the deformation is automatically satisfied due to reparametrization invariance of \eqref{eq:eff0}. Rotations of the normal vectors leave \eqref{eq:eff0} invariant. }
\beq \label{eq:eom}
T^{ab}{K_{ab}}^{i}=2{n_\mu}^{i}\nabla_b\left({\mathcal{S}_a}^{j\mu}{K^{ab}}_j\right)+\mathcal{S}^{akj}{R^{i}}_{akj}~~,
\eeq
\beq \label{eq:bdy}
\eta_a\mathcal{S}^{aij}|_{\partial\mathcal{W}_{p+1}}=0~~,~~\eta_a\left(T^{ab}{e^{\mu}}_b-2{{\mathcal{S}^{b}}_{i}}^{j}{n^{\mu}}_j{K_{b}}^{a i}\right)|_{\partial\mathcal{W}_{p+1}}=0~~,
\eeq
where $R_{\mu\nu\lambda\rho}$ is the Riemann curvature tensor of the background and $\eta_a$ is the normal vector to the submanifold boundary $\partial\mathcal{W}_{p+1}$. We have introduced the worldvolume stress tensor $T^{ab}$ and spin current $\mathcal{S}^{aij}$ which are defined as
\beq
T^{ab}=-\frac{2}{\sqrt{|\gamma|}}\frac{\delta \mathcal{F}}{\delta\gamma_{ab}}~~,~~{\mathcal{S}^{a}}_{ij}=-\frac{1}{\sqrt{|\gamma|}}\frac{\delta \mathcal{F}}{\delta{\omega_a}^{ij}}~~.
\eeq
Particularising to the case of the free energy \eqref{eq:eff0}, the stress tensor and spin current read
\beq
T^{ab}=-\frac{1}{4\widehat\Omega G}\textbf{k}\left(1-\frac{2}{\widehat \Omega}\textbf{k}u^{c}\omega_c\right)\gamma^{ab}+\frac{1}{4\widehat\Omega G}\textbf{k}\left(1-\frac{2}{\widehat \Omega}\textbf{k}u^{c}\omega_c\right)u^{a}u^{b}~~,~~S^{aij}=\frac{1}{4\widehat\Omega^2 G}\textbf{k}^2u^{a}\epsilon^{ij}~~,
\eeq 
in which the stress tensor takes the form of a perfect fluid with pressure $\mathcal{P}$ and energy density $\mathcal{E}$ such that
\beq
\mathcal{P}=-\frac{1}{4\widehat\Omega G}\textbf{k}\left(1-\frac{2}{\widehat \Omega}\textbf{k}u^{c}\omega_c\right)~~,~~\mathcal{E}=-2\mathcal{P}~~.
\eeq
Ignoring the corrections of $\mathcal{O}\left(\varepsilon\right)$ in the effective free energy \eqref{eq:eff0}, which amounts to ignoring the right hand side of \eqref{eq:eom} and the terms involving $S^{aij}$ in \eqref{eq:bdy}, leads to the equation of motion and boundary condition
\beq \label{eq:eom1}
K^{i}=u^{a}u^{b}{K_{ab}}^{i}~~,~~\textbf{k}|_{\partial\mathcal{W}_{p+1}}=0~~,
\eeq
where $K^{i}\equiv \gamma^{ab}{K_{ab}}^{i}$ is the mean extrinsic curvature. The equation of motion above expresses a balance of mechanical forces between brane tension and fluid acceleration in the normal direction while the boundary condition implies that the fluid must move at the speed of light on the boundary.

\subsection{Regime of validity} \label{sec:regval}
The effective theory employed here requires the existence of a hierarchy of scales. There can be multiple hierarchies of scales (or multiple perturbative parameters) depending on the type of deformation applied to the brane: worldvolume, background, bending or spin deformations. As explained in \cite{Armas:2015kra}, for stationary configurations, the length scales associated with each of these deformations at a given order can be determined by the terms appearing in the effective free energy \eqref{eq:eff} at a higher order. At the specific order that we are dealing with here, the validity of the approach considered here requires
\beq \label{eq:val}
r_0\ll\left(|u^{a}\omega_a|^{-1}~,|\mathcal{R}|^{-\frac{1}{2}},|\mathfrak{a}^b\mathfrak{a}_b|^{-\frac{1}{2}}~,~|\omega_{ab}\omega^{ab}|^{-\frac{1}{2}}~,~|K_iK^i|^{-\frac{1}{2}}~,~|R_{||}|^{-\frac{1}{2}}~,~|R_{//}|^{-\frac{1}{2}}\right)~~,
\eeq
where we have defined $R_{||}=\gamma^{ac}\gamma^{bd}R_{abcd}$ and $R_{//}=u^{a}u^{c}\gamma^{bd}R_{abcd}$ and introduced the worldvolume Ricci scalar $\mathcal{R}$ as well as the fluid acceleration $\mathfrak{a}^{b}$ and vorticity $\omega_{ab}$ according to 
\beq
\mathfrak{a}^{b}=u^{a}\nabla_a u^{b}~~,~~\omega_{ab}={P^{c}}_{a}{P^{d}}_{b}\nabla_{[c}u_{d]}~~,
\eeq
where $P_{ab}=\gamma_{ab}+u_{a}u_{b}$ is the projector orthogonal to $u^{a}$. The first condition in \eqref{eq:val} determines the hierarchy of scales due to spin deformations, while the following three determine the hierarchy of scales for worldvolume deformations. The term $|K_iK^i|^{-\frac{1}{2}}$ is the scale associated with bending while the last two in \eqref{eq:val} determine the scales associated with the background curvature. 

The last six terms appearing in \eqref{eq:val} are the higher order corrections that can appear in \eqref{eq:eff} \cite{Armas:2013hsa}. All terms appearing in \eqref{eq:val}, including the coupling to the spin connection, are finite thickness corrections to the brane geometry. As explained in \cite{Armas:2011uf}, these corrections dominate over backreaction or self-force corrections (which are not included in \eqref{eq:eff}) only when $n>2$. Since the case dealt with in this paper is the specific case of $n=1$, one cannot expect, for example, that the corrections due to the spin-orbit coupling in \eqref{eq:eff} will yield the complete result at first order in $\mathcal{O}(\varepsilon)$. While this specific term is important for evaluating the regime of validity, taking it into account in the equilibrium condition can only yield a partial answer. For this reason, we only briefly consider its effect when analysing the solution space of extremal black saturns. We leave the cases of $n>2$ for a future publication \cite{toappear}.

Besides the requirements \eqref{eq:val}, other specific cases impose further restrictions. In particular, if the submanifold has a boundary, say located at $\rho_+$, and $\epsilon$ is the distance away from it, then we must require \cite{Armas:2015kra}
\beq \label{eq:edge}
\rho_+-\epsilon\gg \ell~~,
\eeq
where $\ell$ is the minimum scale that can be probed with the long-wavelength expansion introduced here. This in particular implies that, a priori, submanifolds with boundaries lie outside the regime of validity of this approach. However, there are certain cases for which the method employed here has worked better than expected, such as in the case of Myers-Perry black holes \cite{Emparan:2009vd, Armas:2015kra}. Another example of this will be given in Sec.~\ref{sec:MPX} in which extremal Myers-Perry black holes are shown to be accurately captured by this formalism.

In addition, if configurations are being constructed in backgrounds with other black hole horizons, such as Myers-Perry black hole or black ring horizons, one must require that the distance $d$, between the location of the black hole horizon in the background and the location of the submanifold, satisfies \cite{Emparan:2009vd}
\beq \label{eq:hor}
r_0\ll d~~,
\eeq
such that the interaction between the two horizons can be neglected. 

\subsection{Thermodynamics} \label{sec:thermo}
The free energy \eqref{eq:eff0} allows to extract the thermodynamic properties of the configurations by taking appropriate derivatives with respect to the thermodynamic potentials. In particular, the angular momenta $J_{(l)}$ along the worldvolume directions and the transverse angular momentum $J_\perp$ are given by
\beq \label{eq:js}
\begin{split}
J_{(l)}&=-\frac{\partial\mathcal{F}}{\partial \Omega_{(l)}}|_{T=0}=-\frac{1}{4\widehat\Omega G}\int_{\mathcal{B}_p}\sqrt{|\gamma|}\left[\frac{\partial\textbf{k}}{\partial \Omega_{(l)}}\left(1-\frac{2}{\widehat \Omega}\textbf{k}u^{a}\omega_a\right)-\frac{2}{\widehat \Omega}\frac{\partial \textbf{k}^{a}}{\partial \Omega_{(l)}}\omega_a\right]~~,\\
J_\perp&=-\frac{\partial\mathcal{F}}{\partial \widehat\Omega}|_{T=0}=\frac{1}{4\widehat\Omega^2 G}\int_{\mathcal{B}_p}\sqrt{|\gamma|}\textbf{k}\left(1-\frac{4}{\widehat \Omega}\textbf{k}u^{a}\omega_a\right)~~.
\end{split}
\eeq
The thermodynamic mass can be obtained by using the fact that the free energy, at zero temperature, satisfies the relation
\beq \label{eq:fT}
\mathcal{F}|_{T=0}=M-\sum_{l}\Omega^{(l)}J_{(l)}-\widehat\Omega J_\perp~~.
\eeq
These thermodynamic quantities obey the Smarr relation 
\beq
(D-3)M-(D-2)\left(\sum_{l}\Omega^{(l)}J_{(l)}+\widehat\Omega J_\perp\right)=\hat{\boldsymbol{\mathcal{T}}}~~,
\eeq
where $\hat{\boldsymbol{\mathcal{T}}}$ is the total integrated tension or binding energy. This tension is non-trivial if there are length scales associated with the background spacetime $g_{\mu\nu}$. If the background is Minkowski, the tension vanishes but if the background is, for example, a Schwarzschild black hole with mass parameter $m$, the tension can be obtained using the formula \cite{Armas:2015qsv}
\beq
\hat{\boldsymbol{\mathcal{T}}}=m\frac{\partial \mathcal{F}}{\partial m}|_{\widehat\Omega,\Omega=\text{fixed}}~~.
\eeq
The configurations constructed in this paper are extremal but they still have an associated entropy. This entropy cannot be obtained from the free energy \eqref{eq:eff0} as it does not depend on $T$ but it may be obtained from \eqref{eq:eff} followed by taking the zero temperature limit. We thus find the entropy
\beq
S=-\frac{\partial \mathcal{F}}{\partial T}|_{T\to0}=\frac{\pi}{2\widehat\Omega^2 G}\int_{\mathcal{B}_p}\sqrt{|\gamma|}\textbf{k}\left(1+4\frac{\textbf{k}}{\widehat\Omega}u^{a}\omega_{a}\right)~~.
\eeq
In this work we consider worldvolumes in the background of other black holes, such as rings surrounding a Myers-Perry black hole. In such situations, as explained in \cite{Armas:2015qsv}, the mass $M$ introduced in \eqref{eq:fT} does not correspond to the physical Komar mass measured near the black hole horizon associated with the embedded submanifold. The physical mass must take into account the binding energy and reads \cite{Armas:2015qsv}
\beq
\widehat M=M-\frac{\hat{\boldsymbol{\mathcal{T}}}}{(D-3)}~~.
\eeq
When dealing with such composite systems of two horizons we introduce the total (asymptotic) mass $M^T$, angular momenta $J_{(l)}^{T},J_\perp^{T}$ and entropy $S^{T}$ according to
\beq \label{eq:mass}
M^T=M^{\text{bg}}+\widehat M~~,~~J_{(l)}^{T}=J_{(l)}^{\text{bg}}+J_{(l)}~~,~~J_\perp^{T}=J_\perp^{\text{bg}}+J_\perp~~,~~S^{T}=S^{\text{bg}}+S~~,
\eeq
where $M^{\text{bg}},J_{(l)}^{\text{bg}},J_\perp^{\text{bg}},S^{\text{bg}}$ are the same thermodynamic quantities associated with the background black hole. The fact that the total mass, angular momenta and entropy is a simple addition of the quantities associated with each horizon can be proven in full generality \cite{Elvang:2007hg}.


\section{Extremal black holes in $D=5$} \label{sec:bh5d}
In this section, using the effective theory introduced above, we provide evidence for the existence of many new doubly-spinning extremal black holes with isolated and disconnected horizons in $D=5$ asymptotically flat space. These black holes are ultraspinning in one rotation plane and have finite rotation in the other rotation plane. This allows to, for example, construct extremal doubly-spinning black rings, which we compare against the analytic Pomeransky-Sen'kov extremal black ring solution and find perfect agreement. This is followed by the construction of novel extremal black holes with disconnected horizons, including extremal black saturns, di-rings and bi-rings. At the end of this section, we study the phase structure of extremal black holes in $D=5$.


\subsection{Extremal doubly-spinning black rings} \label{sec:extremaldoubly}
In this section we show that the effective theory described in Sec.~\ref{sec:kerrbrane} accurately reproduces the extremal limit of the exact analytic Pomeransky-Sen'kov black ring solution \cite{Pomeransky:2006bd}. Consider the background metric to be Minkowski space in $D=5$ written in the form
\beq \label{eq:dsads}
ds^2=-dt^2+dr^2+r^2\left(d\theta^2+\sin^2\theta d\psi^2+\cos^2\theta d\chi^2\right)~~,
\eeq
where $0\le r<\infty$ while the angular coordinates satisfy $0\le\theta\le\pi$ and $0\le\psi,\chi\le2\pi$.
In this metric we place a two-dimensional ring geometry of radius $R$ by choosing the embedding map
\beq \label{eq:embed}
t=\tau~~,~~r=R~~,~~\theta=0~~,~~\chi=\phi~~.
\eeq
Setting the ring to rotate with angular velocity $\Omega$ along the $\phi$ direction and with angular velocity $\widehat \Omega$ along the transverse $\psi$ direction leads to the induced metric, surface Killing vector field and transverse Killing vector field
\beq \label{eq:ringds}
\ds=-d\tau^2+R^2d\phi^2~~,~~\ku a\partial_a=\partial_\tau+\Omega\partial_\phi~~,~~\kpu\mu\partial_\mu=\widehat\Omega\partial_\psi~~.
\eeq
The corresponding effective free energy \eqref{eq:eff0} takes the form
\beq
\mathcal{F}[\Omega,\widehat\Omega, R]|_{T=0}=\frac{\pi R}{2G} \frac{\k}{\widehat\Omega}~~,~~\textbf{k}^2=1-\Omega^2R^2~~,
\eeq
and is only composed of the zeroth order part since the normal fundamental one-form $\omega_a$ vanishes for the background \eqref{eq:dsads}. This is expected: since the background is not rotating, there is no spin-orbit coupling. As advertised in Sec.~\ref{sec:kerrbrane}, in the case of single isolated horizons, the variation of this effective free energy leads to the same equilibrium condition as for configurations constructed out of non-spinning branes. For the case at hand, this implies that the equilibrium condition is the same as for singly-spinning black rings \cite{Emparan:2009vd}
\beq \label{eq:ring}
\Omega=\frac{1}{\sqrt{2}R}~~.
\eeq
The only non-trivial scale among those presented in \eqref{eq:val} is the one associated with bending, which implies that $r_0\ll R$. Using \eqref{eq:solrb}, this implies that this construction is valid when $\widehat\Omega R\gg1$. Therefore, we identify the expansion parameter $\varepsilon=(\widehat\Omega R)^{-1}$ for the present case.


\subsubsection{Thermodynamics and comparison with the Pomeransky-Sen'kov solution}
The thermodynamic properties of these doubly-spinning black rings are easily computed using formulae \eqref{eq:js}-\eqref{eq:mass} and take the form
\beq \label{eq:tbf}
\begin{split}
M=\frac{3 \pi  R}{G}r_0 ~~,~~J_\chi=\frac{\sqrt{2}\pi  R^2}{G}r_0~~,~~J_\psi=\frac{\pi  R}{\widehat\Omega G}r_0~~.~~S=\frac{4\sqrt{2} \pi ^2 R}{G}r_0^2~~,~~\boldsymbol{\mathcal{T}}=0~~. 
\end{split}
\eeq
Using the fact that the brane thickness $r_0$, as given in \eqref{eq:solrb}, can be written in terms of the transverse angular velocity such that
\beq \label{eq:thickbf}
r_0=\frac{1}{2\sqrt{2}\widehat\Omega}~~,
\eeq
the thermodynamic quantities \eqref{eq:tbf} can all be expressed in terms of the radius $R$ and the transverse angular velocity $\widehat\Omega$.
In order to describe the exact analytic solution, these quantities should reproduce the Pomeransky-Sen'kov extremal black ring solution. Using the Pomeransky-Sen'kov solution as written in \cite{Pomeransky:2006bd}, performing the following redefinitions
\beq
\lambda=\frac{\mu}{R}~~,~~\nu=\frac{a^2}{R^2}~~,~~k=\frac{R}{\sqrt{2}}~~,
\eeq
choosing the extremality condition $a=\mu/2$ and taking the limit $R\to\infty$, the thermodynamic properties obtained in \cite{Elvang:2007hs} become\footnote{Note that the angular coordinates $(\phi,\psi)$ in \cite{Elvang:2007hs} are $(\psi,\phi)$ in \eqref{eq:tanal}. }
\beq \label{eq:tanal}
\begin{split}
&\Omega_\chi=\frac{1}{\sqrt{2}R}~~,~~\Omega_\psi=\frac{1}{\sqrt{2}\mu}~~,~~M=\frac{3\pi R}{2G}\mu~~,~~J_\chi=\frac{\pi R^2}{\sqrt{2}G}\mu~~,\\
&J_\psi=\frac{\pi R}{\sqrt{2}G}\mu^2~~,~~S=\frac{\sqrt{2}\pi^2 R}{G} \mu^2~~,
\end{split}
\eeq
where we have ignored higher order corrections in $\mu/R$ (equivalently, corrections in higher powers of $\varepsilon$), with $\mu$ being the mass parameter and $a$ the spin parameter of the doubly-spinning black ring. The analytic results \eqref{eq:tanal} from the limit of the exact solution are in perfect agreement with \eqref{eq:tbf}-\eqref{eq:thickbf} once one identifies $\Omega_\chi=\Omega$ and $\Omega_\psi=\widehat\Omega$. We note that the expressions for the black hole mass and angular momentum along the $\psi$ direction are actually exact, i.e. they do not receive corrections in $\widehat\Omega R$. In Sec.~\ref{sec:phase} we will compare the phase diagram obtained using \eqref{eq:tbf} with that of the exact solution and show that there is good agreement up to values of $\widehat\Omega R\sim 0.1$. This provides strong evidence for the fact that the effective theory of extremal Kerr branes reproduce the correct analytic results.


\subsection{Extremal black saturns} \label{sec:bs}
In order to build extremal black saturns we consider the background geometry to be that of a Myers-Perry black hole \cite{Myers:1986un} with metric (see e.g. \cite{Harmark:2004rm})
\beq \label{eq:dsMP}
\begin{split}
ds^2=&-dt^2+\frac{m^2}{\Sigma}\left[dt-b\sin^2\theta d\psi-a\cos^2\theta d\chi\right]^2+\frac{\Sigma}{\Delta}dr^2+\Sigma d\theta^2 \\
&+(r^2+b^2)\sin^2\theta d\psi^2+(r^2+a^2)\cos^2\theta d\chi^2~~,
\end{split}
\eeq
where we have introduced the two functions
\beq
\Sigma=r^2+b^2\cos^2\theta+a^2\sin^2\theta~~,~~\Delta=r^2\left(1+\frac{b^2}{r^2}\right)\left(1+\frac{a^2}{r^2}\right)-m^2~~.
\eeq
The parameters $a,b$ are the rotation parameters in the two rotation planes of the Myers-Perry black hole and lie within the range $-\infty<a,b<\infty$.  We are interested in the extremal limit of the metric \eqref{eq:dsMP} for which there is a regular horizon. This is achieved when 
\beq \label{eq:extc}
\{(m^2+a^2+b^2)^2-4a^2b^2=0~\wedge ~a,b\ne0\}~~,
\eeq 
for which the horizon location at $r=r_+$, such that $\Delta(r_+)=0$, is given by
\beq \label{eq:mphor}
r_+=\sqrt{|a|m-a^2}~~.
\eeq
The reality condition on $r_+$, together with the extremality condition \eqref{eq:extc}, implies that the rotation parameters satisfy $-1<\a,\b<1$, where we have defined the dimensionless quantities $\a=a/m$ and $\b=b/m$. At extremality, the angular velocities of the Myers-Perry black hole reduce to
\beq \label{eq:angMP}
\Omega_\psi=\pm \frac{1}{m}~~,~~\Omega_\chi=\pm \frac{1}{m}~~,
\eeq
where the sign of $\Omega_\psi$ is given by the sign of $b$ and the sign of $\Omega_\chi$ by the sign of $a$. This naturally splits the solution space of black saturns in four different regions depending on the signs of the rotation parameters: $\{\Omega_\psi,\Omega_\chi>0$ and $0<\a,\b<1\}$, $\{\Omega_\psi>0$, $\Omega_\chi<0$ and $0<\b<1,-1<\a<0\}$, $\{\Omega_\psi<0$, $\Omega_\chi>0$ and $-1<\b<0,0<\a<1\}$, $\{\Omega_\psi,\Omega_\chi<0$ and $-1<\a,\b<0\}$. 
The angular velocities \eqref{eq:angMP} only depend on the mass parameter, while the angular momenta depend on the rotation parameters (see e.g. \cite{Harmark:2004rm})
\beq
J_\psi=\frac{\pi}{4 G}b m^2~~,~~J_\chi=\frac{\pi}{4 G}a m^2~~,
\eeq
and therefore the magnitude of the rotation parameters will affect the ring configuration that we will now consider.

\subsubsection{Black saturn embedding}
In the extremal geometry \eqref{eq:dsMP} we embed a ring by choosing the same embedding map as in \eqref{eq:embed}. The induced metric is that of a rotating worldsheet
\beq
\ds=-\left(1-\frac{m^2}{R^2+b^2}\right)d\tau^2-2a\frac{m^2}{R^2+b^2}d\tau d\phi+\left(R^2+a^2+\frac{a^2 m^2}{R^2+b^2}\right)d\phi^2~~,
\eeq
while the worldvolume and transverse Killing vectors fields take the same form as in \eqref{eq:ringds}. The modulus of the worldvolume Killing vector field and the determinant of the induced metric take the form
\beq \label{eq:kgsaturn}
\k^2=1-\Omega ^2\left(R^2+a^2\right)-\frac{m^2 (1-a\Omega)^2}{R^2+b^2}~~,~~\gamma=-a^2-R^2 \left(1-\frac{m^2}{R^2+b^2}\right)~~.
\eeq
Since \eqref{eq:kgsaturn} only depends on the sign of $\a$ but not on the sign of $\b$, when introducing \eqref{eq:kgsaturn} into the effective free energy \eqref{eq:eff0}, and ignoring the higher order correction, the resulting equilibrium condition will be insensitive to the sign of $\b$. This implies that at zeroth order, the four regions of solution space defined above collapse into two regions only $\{\Omega_\chi>0~,~0<\a<1\}$ and $\{\Omega_\chi<0~,~-1<\a<0\}$. However, these two regions exhibit reflection symmetry, i.e. sending $\a\to-\a$ and $\Omega\to-\Omega$ leaves \eqref{eq:kgsaturn} unchanged. In order for the configuration to be sensitive to the sign of $\b$ (or equivalent to the sign of $\Omega_\psi$), one needs to take into account the spin-orbit coupling in \eqref{eq:eff0}. We will first consider the zeroth order case, starting off with the assumption of equal rotation parameters, and at the end of this section consider the effect due to the first order correction.

\subsubsection{Equal rotation parameters} \label{sec:bsequal}
In order to gain intuition, it is useful to consider the simplest case in which the centre Myers-Perry black hole is rotating with equal angular momenta. In this case $\a=\b=1/2$, since the extremality condition \eqref{eq:extc} gives $\a+\b=1$, with $\Omega_\psi,\Omega_\chi>0$. There are two possible equilibrium conditions, describing clockwise '+' or anti-clockwise '-' rotation
\beq \label{eq:oequal}
\Omega=\frac{1}{R}\frac{16\m^5\pm\sqrt{2}|\m^2-4|\sqrt{(4+\m^2)(16+32\m^2+11\m^4)}}{(4+3\m^2)(16+3\m^4)}~~,~~\m=\frac{m}{R}~~.
\eeq
However, not all radii $R$ are allowed since we must have that $\k^2,\gamma>0$. Using \eqref{eq:kgsaturn}, this implies that $\m>2$ for the '+' branch and $\m^{-1}>\sqrt{\frac{5}{4}+\sqrt{2}}$ for the '-' branch. This means that the solution in the '+' branch exists for any value of the radius up to the Myers-Perry horizon \eqref{eq:mphor} located at $\m=2$. In general one should not expect solutions with $\m\sim2$ to be valid since they can violate the requirement \eqref{eq:hor} but we will provide a detailed analysis of the regime of validity in the next section.

\begin{figure}[h!] 
\centering
  \includegraphics[width=0.5\linewidth]{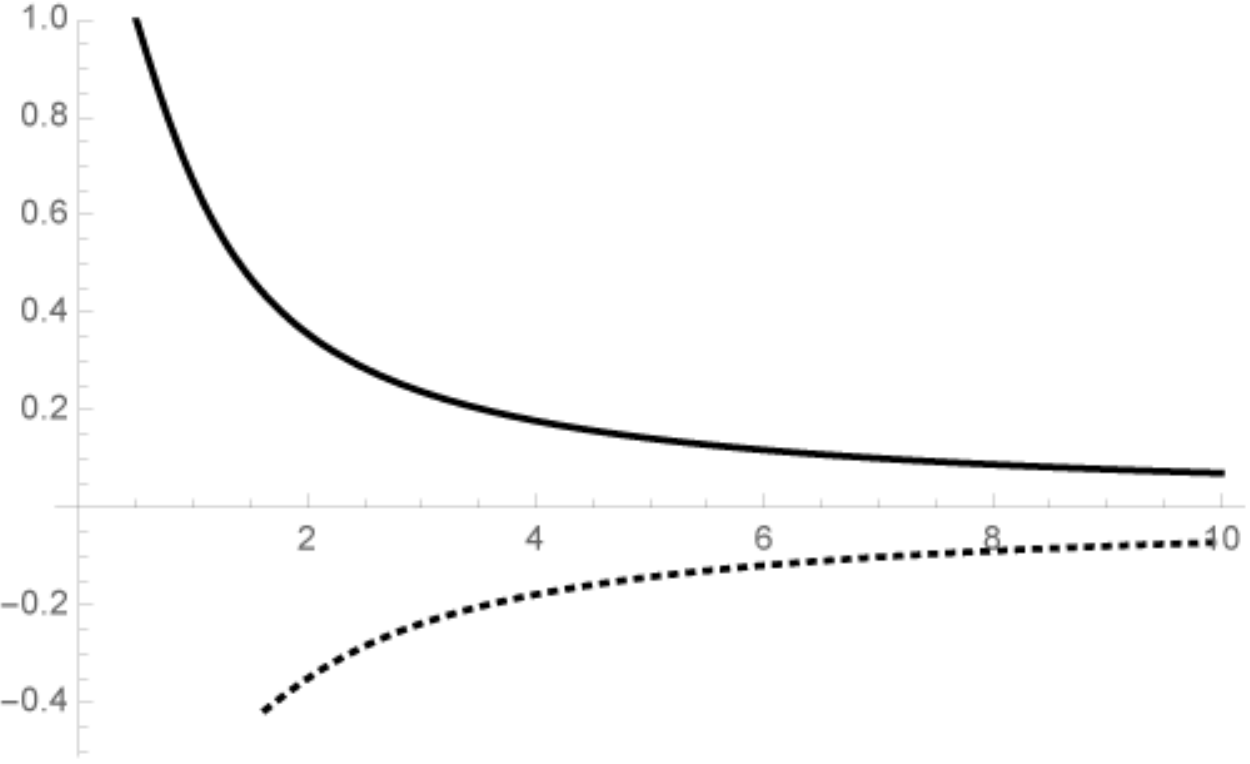}
  \begin{picture}(0,0)(0,0)
  \put(-240,120){ $\Omega $}
    \put(-20,25){ $R$}
\end{picture} 
\caption{$\Omega$ as a function of $R$ in units $m=1$. The black line is the '+' branch while the dashed line is the '-' branch.} \label{fig:BSequal}
\end{figure}
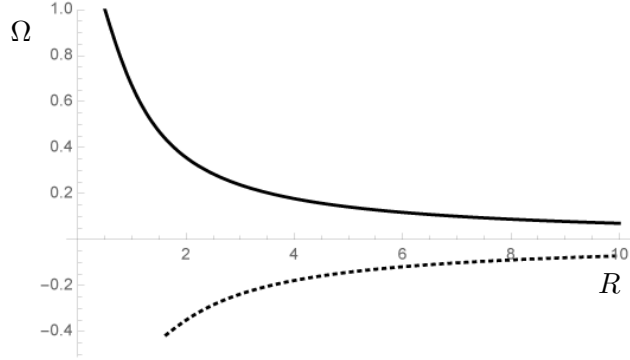

Note that when the centre black hole is removed ($\m=0$) we recover \eqref{eq:ring} with an overall plus or minus sign. The two branches of solutions are qualitatively different as can be seen in Fig.~\ref{fig:BSequal}. 
The '+' branch leads to extremal black saturn solutions for which the ring is rotating along the $\mathbb{S}^1$ in the same (clockwise) direction as the centre Myers-Perry black hole. As the radius of the ring is increased and the gravitational force from the centre black hole is negligible, i.e. in the limit $\m\to0$, the angular velocity of the ring approaches that of the Pomeransky-Sen'kov solution. As one approaches smaller radii, the ring velocity approaches that of the speed of the light since $\k\to0$ as $\m\to2$. This is a consequence of the fact that the gravitational pull of the centre black hole increases as the distance between the ring and the black hole decreases. The '-' branch consists of rings that are rotating counter-clockwise with respect to the centre black hole. In the limit $\m\to0$, the ring velocity approaches  that of the Pomeransky-Sen'kov solution but with an overall negative sign. The '-' branch has lower angular velocity than the '+' branch for any value of the radius. However, the velocity $\k$ of any point on the ring in the '-' branch is higher than in the '+' branch. Since the ring is rotating counter-clockwise compared to the centre black hole, there is an additional contribution to the velocity of each point on the ring compared to the '+' branch. This implies that the speed of light will be reached for larger radii than in the case of the '+' branch, thereby explaining why these two branches do not exhibit reflection symmetry.

\subsubsection{General solution for arbitrary rotation parameter}
We now consider the general solution for an extremal ring surrounding an extremal Myers-Perry black hole with $\a+\b=1$. As in the previous case, the equilibrium condition has two branches of solutions, one which rotates clockwise and another anti-clockwise with respect to the centre black hole. This is given by the expression
\beq \label{eq:eqgenBS}
\Omega=\frac{1}{R}\frac{2 \a\b \m^3\pm\left|1+ \a^2-\m \a\right|\sqrt{\left(1+\b^2\right) \left(\b^2 \left(2 \a^2-8 \a \m+9 \m^2\right)+ (2 \a-3 \m)^2+2\right)}}{g(\m,\a,\b)}~~,
\eeq
where we have defined the function
\beq
\begin{split}
g(\m,\a,\b)=&~2+\left(6 \a^2-10 \a \m+5\m^2\right)- \b \left(6 \a^3-14 \a^2 \m+10 \a \m^2-3 \m^3\right) \\
&-\a \b \left(2 \a^4-8 \a^3 \m+11\a^2 \m^2-8 \a \m^3+\m^4\right)~~.
\end{split}
\eeq
If $\a=\b=1/2$ then \eqref{eq:eqgenBS} reduces to \eqref{eq:kgsaturn}. The black hole rotation parameters $\a,\b$ can be either positive or negative leading to the four different regions of phase space, though as explained above, at zeroth order, these regions collapse into two regions, which are themselves equivalent by an appropriate reflection. 

\begin{figure}[h!] 
\centering
\begin{subfigure}{.5\textwidth}
  \centering
  \includegraphics[width=0.7\linewidth]{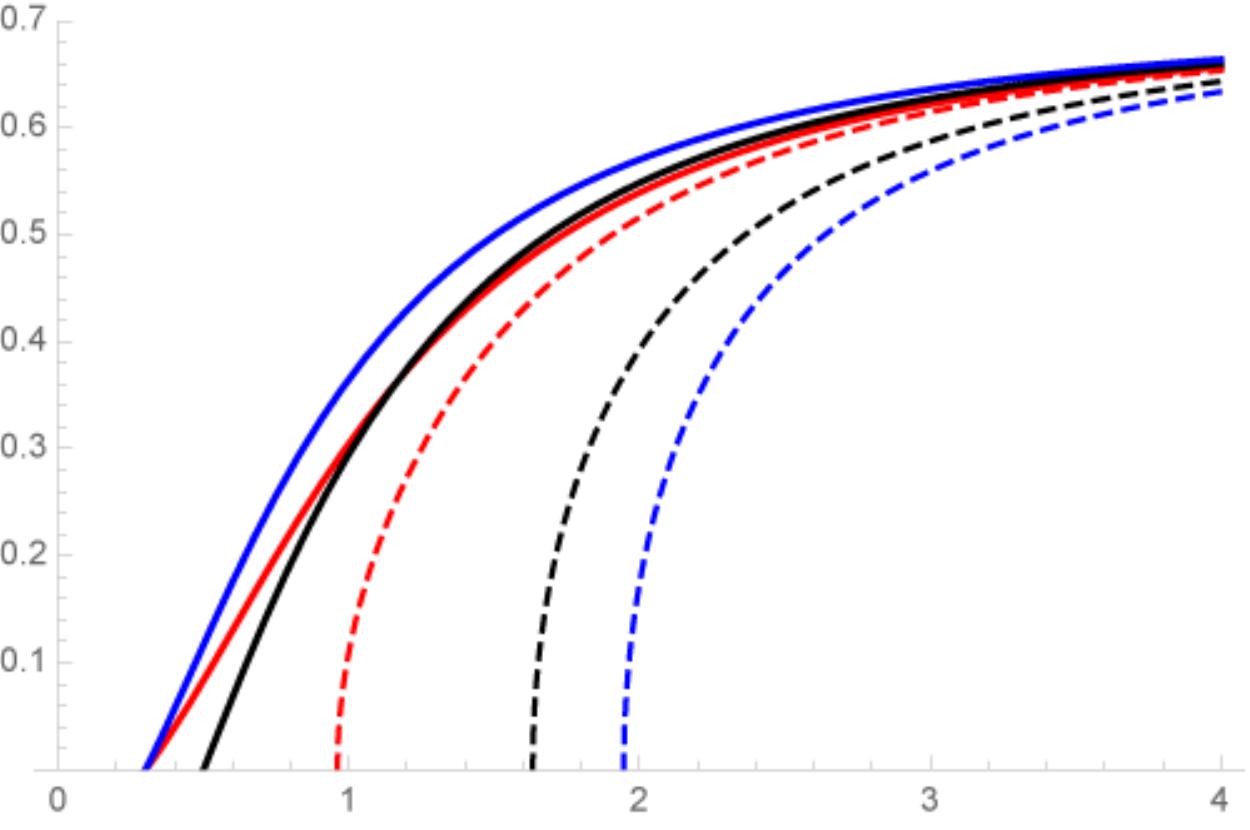}
  \begin{picture}(0,0)(0,0)
  \put(-175,85){ $\k $}
    \put(-25,-5){ $R$}
\end{picture}  
\end{subfigure}%
\begin{subfigure}{.5\textwidth}
  \centering
  \includegraphics[width=0.7\linewidth]{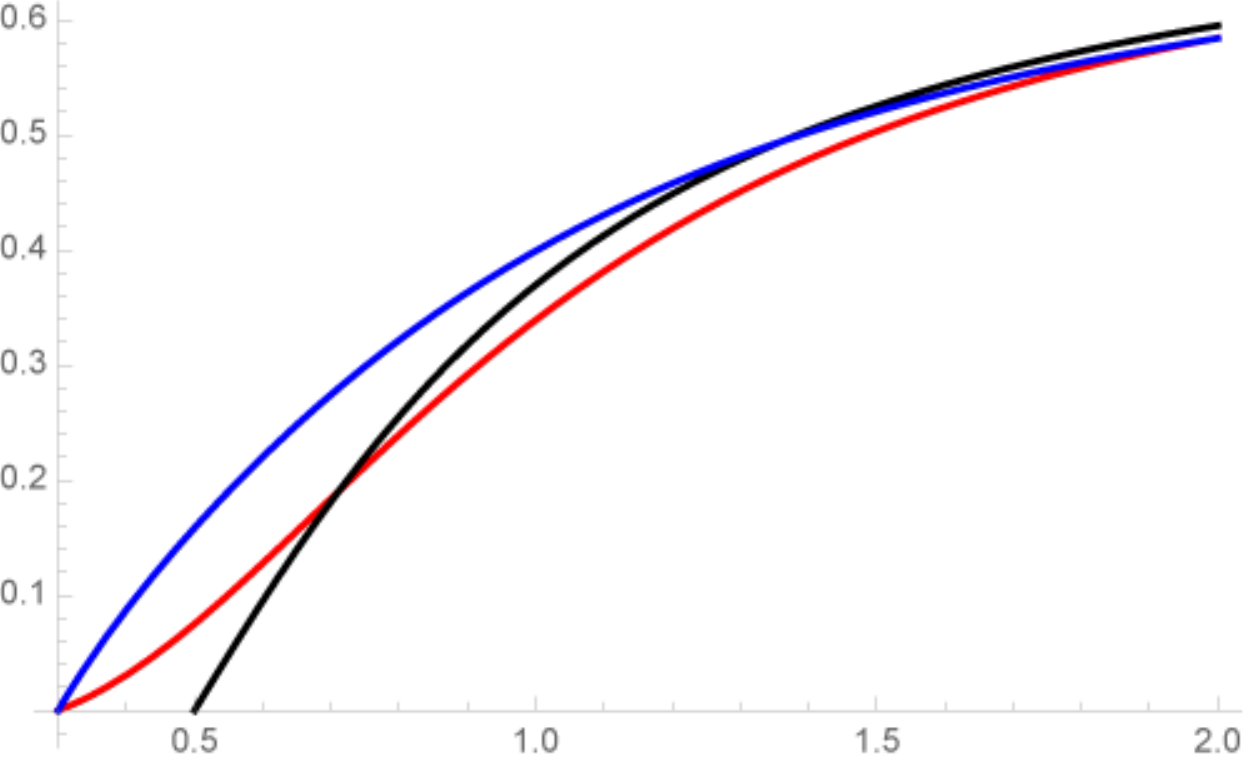}
  \begin{picture}(0,0)(0,0)
  \put(-194,85){ $\frac{\partial\sqrt{-\gamma}}{\partial R} \k$}
    \put(-25,-5){ $R$}
\end{picture} 
\end{subfigure}%
\caption{On the l.h.s. the ring velocity $\k$ is shown as a function of $R$ in units $m=1$ for the values of the rotation parameter $a=1/10$ (red), $a=1/2$ (black) and $a=9/10$ (blue) while on the r.h.s.  $\frac{\partial\gamma}{\partial R} \k$ is shown as a function of $R$ for the same values of $a$ in the '+' branch. On the l.h.s., the solid lines correspond to the '+' branch while the dashed lines correspond to the '-' branch. The black line corresponds to the case of Fig.~\ref{fig:BSequal}.}  
  \label{fig:BSgen}
\end{figure}

\paragraph{Region $\Omega_\chi>0$ and $0<\a,\b<1$:} In this case the centre black hole is rotating in the clockwise direction along the $\chi$ direction. In the limit $\m\to0$, \eqref{eq:eqgenBS} leads to the equilibrium condition for the Pomeransky-Sen'kov solution in spheroidal coordinates $\Omega^{-1}=\sqrt{2}\sqrt{R^2+a^2}$. In order to compare the same features between different values of the rotation parameter, in Fig.~\ref{fig:BSgen} we depict the velocity of the ring $\k$ (which has the limit $\k\to1/\sqrt{2}$ as $\m\to0$) as a function of the radius $R$ for different values of the rotation parameter as well as $(\partial\sqrt{-\gamma}/\partial R)~\k$, which is a measure of the gravitational force acting on the worldsheet of the ring due to the centre Myers-Perry black hole.
The figure on the r.h.s. exhibits the strength of the gravitational field acting on the ring due to the centre black hole for three different values of the rotation parameter $a$. For values of $R\gtrsim1$, the gravitational field strength increases with increasing rotation parameter. However, for values $R\lesssim1$ this is not necessarily the case and in fact for $a=1/10$ the gravitational field strength is higher than for $a=1/2$. This leaves an imprint in the ring velocity $\k$. In the '-' branch, due to the counter-clockwise rotation of the ring, the speed of light ($\k=0$) is approached for higher values of $R$ compared to the '+' branch as in the previous case of $\a=\b=1/2$. However, since in the '-' branch these values of $R$ are such that $R\gtrsim1$, the higher the rotation parameter, the higher the velocity of the ring must be for a given value of the radius. This is in contrast with the '+' branch for which the speed of light is approached for lower values of $R$. Since this happens for $R\lesssim1$, the ring rotates with higher velocity for a given value of the radius for $a=1/10$ than for $a=1/2$. 

The condition $\gamma=0$ dictates the minimum radius that can be approached in the '+' branch and gives $R=r_+$, which is when the ring is placed at the horizon \eqref{eq:mphor} and the solution is not valid. This, however, implies that $\gamma$ has the same zero for  $a=1/10$ and $a=9/10$ and explains why the minimum radius in the '+' branch is the same for these two values of the rotation parameter.

\subsubsection{Validity analysis} 
We have presented above the solution space for black saturns but have not inquired about its regime of validity. The non-trivial scales in \eqref{eq:val} are those associated with the background curvature, spin and bending deformations. From the scale associated with bending deformations we find that we must have
\beq \label{eq:vc1}
\widehat\Omega \ell \gg 1~~,
\eeq
where $\ell\equiv\text{min}(R,m)$. The two background curvature scales in \eqref{eq:val} require \eqref{eq:vc1} with $\ell=m$ . Evaluating the scale associated with spin deformations
\beq \label{eq:spinscale}
|u^{a}\omega_a|^{-1}=\textbf{k}\frac{(b^2+R^2)^2}{bm^2(1-a\Omega)}~~,
\eeq
leads to $\widehat\Omega m\gg1$ for $R\to r_+$, which is guaranteed by \eqref{eq:vc1}.

The validity condition \eqref{eq:hor} requires the two horizons to be sufficiently far apart, therefore we evaluate the ratio $|(R-r_0)-r_+|/r_0$. Even for distances very close to the centre black hole $R\sim r_+$ one can satisfy the requirement $r_0\ll d$ in the '+' branch as long as $\widehat\Omega m\gg \sqrt{\a}$, which is guaranteed by \eqref{eq:vc1}. Note that no such requirement is imposed in the '-' branch, as the ring velocity reaches the speed of light for larger ring radii. Given these considerations, we identify the small perturbative parameter $\varepsilon\equiv(\widehat\Omega m)^{-1}\sim r_0/m$.

\subsubsection{The spin-orbit coupling}
We now briefly study the effect of considering the coupling to the spin connection in the free energy \eqref{eq:eff0}. Using \eqref{eq:spinscale}, the free energy can be written as
\beq \label{eq:effc}
\mathcal{F}[\widehat\Omega,\Omega, R,g_{\mu\nu}]|_{T=0}=\frac{1}{4\widehat\Omega G}\int_{\mathcal{B}_p}\sqrt{|\gamma|}\textbf{k}\left(1-\frac{2\varepsilon \b}{(1+\m^{-2})}(1-a\Omega)\right)~~.
\eeq
Clearly, if the background spacetime is not rotating $\b=0$, the spin-orbit coupling vanishes. As mentioned in Sec.~\ref{sec:regval}, the corrections appearing in the effective free energy \eqref{eq:eff0} for $n\le2$ may be subleading compared to backreaction corrections. In the present case, backreaction corrections will enter at the same order as the correction due to spin-orbit effects. Therefore, while we briefly study the effect of such coupling in this section, one cannot trust this to be the full result at order $\mathcal{O}(\varepsilon)$.

First note that due to \eqref{eq:kgsaturn}, the free energy \eqref{eq:effc} is invariant under the shifts $a\to-a$ and $\Omega\to-\Omega$. Therefore one is free to take $a>0$ without loss of generality. This means that the four regions introduced above collapse into two. In addition, since we have chosen $\widehat\Omega>0$, the free energy \eqref{eq:effc} is not invariant under the shift $\b\to-\b$, leading to two different regions of solution space depending on the sign of $\b$. We can solve the equilibrium condition that arises from varying \eqref{eq:effc} with respect to $R$ (i.e. \eqref{eq:eom}) by making the perturbative ansatz
\beq
\Omega=\Omega_0+\varepsilon~\Omega_1~~,
\eeq
where for simplicity we have chosen $\a=\b=1/2$, for which $\Omega_0$ is the zeroth order result \eqref{eq:oequal}. Using this we may determine $\Omega_1$ in the '+' branch to be
\beq
\Omega_1 R=\frac{2\sqrt{2}}{8+25\m^2}\m^2+\mathcal{O}\left(\m^3\right)~~,
\eeq
for small $\m$. In this region for which $\b>0$, the angular velocity increases due to the spin-orbit coupling. However, this result is only part of the full first order correction to the angular velocity of the ring. 


\subsection{Extremal di-rings} \label{sec:diring}
Di-rings are black hole solutions with two concentric disconnected ring horizons with topology $\left(\mathbb{S}^{1}\times \mathbb{S}^{2}\right)\cup\left(\mathbb{S}^{1}\times \mathbb{S}^{2}\right)$ \cite{Evslin:2007fv, Iguchi:2007is}. These analytic solutions have finite temperature and no regular zero temperature limit. Each of the rings is rotating only along the $\mathbb{S}^{1}$ direction. On the other hand, extremal di-rings can be considered by placing a ring geometry around an extremal doubly-spinning black ring. The Pomeransky-Sen'kov solution \cite{Pomeransky:2006bd} has metric \footnote{In order to match the conventions of \cite{Pomeransky:2006bd} one must relabel $\widehat\omega(x,y)\to\Omega$, $\chi\to\phi$, $R_\circ^2\to2 k^2$ and change Minkowski metric sign conventions to $\eta_{\mu\nu}=\text{Diag}(+1,-1,-1,-1,-1)$.}
\beq\label{eq:dsPS}
\begin{split}
ds^2=&-\frac{H(y,x)}{H(x,y)}(dt+\widehat\omega(x,y))^2-\frac{F(x,y)}{H(y,x)}d\chi^2-2\frac{J(x,y)}{H(y,x)}d\chi d\psi+\frac{F(y,x)}{H(y,x)}d\psi^2 \\
&+\frac{R_\circ^2H(x,y)}{(x-y)^2(1-\nu)^2}\left(\frac{dx^2}{G(x)}-\frac{dy^2}{G(y)}\right)~~,
\end{split}
\eeq
where the function $G(y)=(1-y^2)(1+\lambda y+\nu y^2)$ determines the locations of the inner and outer horizons and the coordinate $y$ has the range $-\infty<y<-1$. The parameters $\lambda,\nu$ are dimensionless while $R_\circ$ is a dimensionful parameter determining the radius of the ring. We are interested in the extremal limit for which $\lambda=2\sqrt{\nu}$ with $0<\nu<1$. In this case the extremal horizon is located at $y_\text{h}=-\nu^{-\frac{1}{2}}$ and the remaining functions introduced in \eqref{eq:dsPS} are defined in \cite{Pomeransky:2006bd}.  Using the thermodynamic properties of the Pomeransky-Sen'kov solution obtained in \cite{Elvang:2007hs}, one observes that at extremality $\lambda=2\sqrt{\nu}$ both the angular momenta of the extremal doubly-spinning black ring satisfy $J_\psi, J_\chi>0$ while the angular velocities read
\beq \label{eq:omegabr}
\Omega_\psi=\frac{1}{2\sqrt{2}R_\circ}\left(\frac{(1-\sqrt{\nu})(1+\nu)}{\nu+\sqrt{\nu}}\right)~~,~~\Omega_\chi=\frac{1}{\sqrt{2}R_\circ}\left(\frac{1-\sqrt{\nu}}{1+\sqrt{\nu}}\right)~~,
\eeq
and satisfy $\Omega_\psi,\Omega_\chi>0$. This implies that the solution space of di-rings is comparatively simpler than the case of the black saturn.

\subsubsection{Di-ring embedding}
In order to better understand how to embed a second (outer) extremal ring around the (inner) ring geometry \eqref{eq:dsPS} it is useful to consider a change of coordinates that makes the metric \eqref{eq:dsPS} manifestly asymptotically flat at large distances away from the ring horizon. This can be accomplished by performing the coordinate transformation \cite{Durkee:2008an}
\beq \label{eq:rt}
x=-1+\frac{2R^2_\circ}{r^2}\frac{1+\nu-\lambda}{1-\nu}\cos^2\theta~~,~~y=-1-\frac{2R_\circ^2}{r^2}\frac{1+\nu-\lambda}{1-\nu}\sin^2\theta~~,
\eeq
such that $R_\circ\sqrt{(1+\nu-\lambda)/(1-\nu)}\le r<\infty$ and $0\le\theta\le\pi$. In this case, asymptotic infinity is reached when $r\to\infty$ and the metric \eqref{eq:dsPS} becomes
\beq
ds^2\approx-dt^2+dr^2+r^2\left(d\theta^2+\cos^2\theta d\psi^2+\sin^2\theta d\chi^2\right)~~.
\eeq
In the background \eqref{eq:dsPS} but with coordinates $(x,y)\to(r,\theta)$ we want to place a concentric ring rotating along the $\chi$ direction. This can be done by setting $\theta=\pi/2$ and $r=R$ which in turn leads to\footnote{The fact that one must set $x=-1$ and $y=\text{constant}$ can be understood directly by working with $(x,y)$ coordinates since the geometry of such cross sections are circles in the $\psi$ plane of rotation \cite{Emparan:2006mm}.}
\beq
x=-1~~,~~y=y_{R}=-1-\frac{2R_\circ^2}{R^2}\frac{1+\nu-2\sqrt{\nu}}{1-\nu}~~,
\eeq
and choosing the embedding coordinates $t=\tau~,~\chi=\phi$. The constant $y_R$ lies in the interval $y_\text{h}<y_R<-1$. Introducing this into \eqref{eq:dsPS} leads to the induced metric 
\beq
\ds=-\frac{H(y_R,-1)}{H(-1,y_R)}(d\tau+\widehat\omega(-1,y_R))^2-\frac{F(-1,y_R)}{H(y_R,-1)}d\phi^2~~.
\eeq
The functions $H(-1, y_R)$, $H(y_R,-1)$, $F(-1,y_R)$ and the one form $\widehat{\omega}(-1,y_R)$ are given by
\beq \label{eq:brfunctions}
\begin{split}
H(-1,y_R)=&-\nu ^2+4 \nu +4 \sqrt{\nu } \left(\nu ^2 y_R^2-1\right)-\nu  (\nu  (\nu +4)-1) y_R^2+1~~,\\
H(y_R,-1)=&-\nu ^2+4 \nu +4 \nu ^{3/2} \left(y_R^2-1\right)-\nu  (\nu  (\nu +4)-1) y_R^2-4 \sqrt{\nu }\left(\nu ^2-1\right) y_R+1~~,\\
F(-1,y_R)=&-\frac{R_\circ^2 \left(\sqrt{\nu }-1\right)^4 (y_R-1) \left(\sqrt{\nu } y_R+1\right)^2}{y_R+1}~~,\\
\widehat\omega(-1,y_R)=&-\frac{4 R_\circ \sqrt{(\nu -1)^2} \sqrt{\nu } (y_R+1) \left(3 \sqrt{\nu }+\nu
   ^{3/2} y_R+3 \nu  y_R+1\right)}{\sqrt{2}\left(\sqrt{\nu }-1\right)^2 u(\nu,y_R)}d\phi~~,\\
u(\nu,y_R)=&\sqrt{\nu } \left(\nu+5 \sqrt{\nu }+\left(\nu ^{3/2}+\nu ^2+5 \nu +\sqrt{\nu }\right) y_R^2+4 \left(\sqrt{\nu}+1\right) (\nu +1) y_R+1\right)+1 ~~.
\end{split}
\eeq
The worldvolume and transverse Killing vector fields are taken to be of the form \eqref{eq:ringds}. The modulus of the worldvolume Killing vector field and the determinant of the induced metric read
\beq \label{eq:kgdiring}
\k^2=-\frac{F(-1,y_R)H(-1,y_R)\Omega^2+H(y_R,-1)^2\left(1+\widehat\omega(-1,y_R)\Omega\right)^2 }{H(-1,y_R)H(y_R,-1)}~~,~~\gamma=-\frac{F(-1,y_R)}{H(-1,y_R)}~~.
\eeq
We will now examine the space of possible solutions.
\subsubsection{Solution space}
Introducing \eqref{eq:kgdiring} into the free energy \eqref{eq:eff0}, ignoring higher order corrections, and varying it with respect to $R$ leads to the equation for mechanical equilibrium. This equation admits two branches of solutions as in the case of the black saturn. The '+' branch is rotating in the same direction along the $\mathbb{S}^{1}$ as the centre black ring while the '-' is rotating in the counter-clockwise direction. These solutions exhibit a cumbersome relation between the angular velocity and the ring radius $R$ but can nevertheless be determined analytically. For clarity of presentation, below we present the analytic result for small $\textbf{R}_\circ\equiv R_\circ/R$
\beq \label{eq:Omegaexp}
\Omega=\pm\frac{1}{\sqrt{2}R}\left(1+\frac{2\sqrt{\nu}-8\nu-2\nu^{2/3}+\nu^2-1}{2(1-\nu)^2}\textbf{R}_\circ^2+\mathcal{O}\left(\textbf{R}_\circ^4\right)\right)~~.
\eeq
The zeroth order behaviour is the expected behaviour of an isolated ring horizon in flat space time, i.e. \eqref{eq:ring}. The two branches have the same behaviour (modulo a minus sign) at small $\textbf{R}_\circ$ but this does not continue to be the case once further corrections are considered.

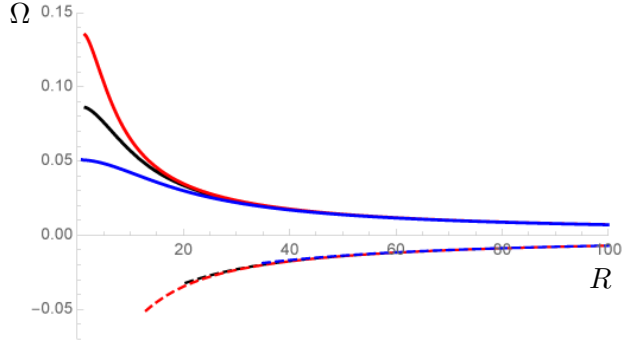
\begin{figure}[h!] 
\centering
  \includegraphics[width=0.5\linewidth]{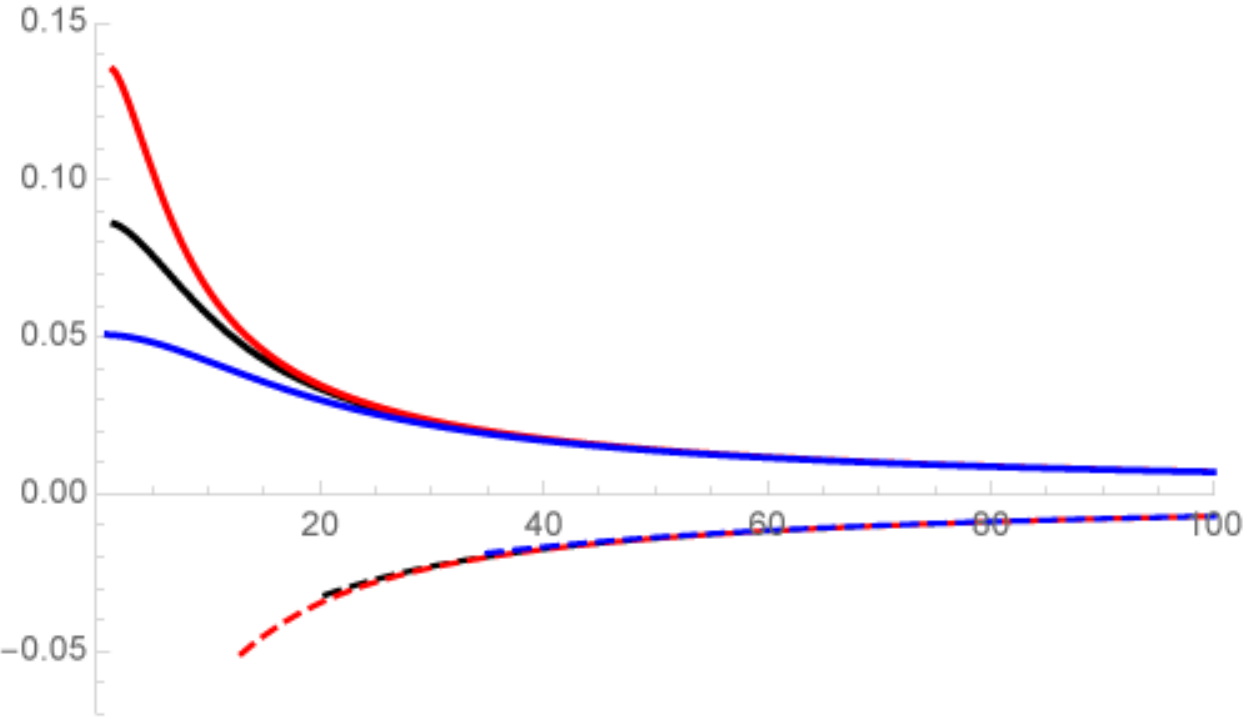}
  \begin{picture}(0,0)(0,0)
  \put(-237,120){ $\Omega $}
    \put(-20,20){ $R$}
\end{picture} 
\caption{$\Omega$ as a function of $R$ in units $R_\circ=1$. The upper thick curves are the '+' branch of solutions rotating in the same direction as the centre black ring while the lower '-' branch of dashed curves is rotating in the opposite direction. These curves represent three values of $\nu$: $\nu=1/3$ (red), $\nu=1/2$ (black) and $\nu=2/3$ (blue).} \label{fig:diring}
\end{figure}

In Fig.~\ref{fig:diring} we show the behaviour of the angular velocity as a function of $R$, which is valid for $\widehat\Omega \ell\gg1$ with $\ell\equiv\text{min}(R,R_\circ)$. As the ring radius $R$ is increased, its angular velocity decreases. This is the same behaviour as found for all ring geometries considered previously and it is rooted in the fact that as the ring size $R$ increases, the gravitational field due to the inner black ring decreases in strength.
The behaviour of the solution space is qualitatively the same as in the region $\Omega_\psi,\Omega_\chi>0$ and $0<\a,\b<1$ of the black saturn solution. The curves terminate at values of $R$ where the ring velocity $\k$ reaches the speed of light, which happens when the outer ring sits at the horizon of the inner ring and the solution breaks down. In the '+' branch the minimum  radius decreases with increasing $\nu$ while in the '-' branch the opposite behaviour is observed. This is due to the fact that the angular momentum $J_\psi$ of the centre black ring increases with increasing $\nu$. Furthermore, no static solutions are possible.


\subsection{Extremal bi-rings} \label{sec:biring}
An analytic solution for non-extremal bi-rings was obtained in \cite{Izumi:2007qx, Elvang:2007hs} and has topology $\left(\mathbb{S}^{1}\times \mathbb{S}^{2}\right)\cup\left(\mathbb{S}^{1}\times \mathbb{S}^{2}\right)$. Contrary to di-rings,  it consists of two non-concentric rings in which the $\mathbb{S}^{1}$ direction of each ring lies on two different orthogonal planes. The analytic bi-ring solution has non-zero angular momentum only along each of the $\mathbb{S}^{1}$ directions. Here we present evidence for the existence of extremal bi-ring configurations with angular momentum also in the $\mathbb{S}^{2}$.

\subsubsection{Bi-ring embedding}
In order to construct extremal bi-ring solutions, one embeds the ring geometry in the background \eqref{eq:dsPS} in an orthogonal plane to the $\mathbb{S}^{1}$ of the centre black ring. This can be done by using the $(r,\theta)$ coordinates \eqref{eq:rt} and choosing $\theta=0$ and $r=R$ such that
\beq
x=x_R=-1+\frac{2R^2_\circ}{R^2}\frac{1+\nu-2\sqrt{\nu}}{1-\nu}~~,~~y=-1~~,~~t=\tau~~,~~\psi=\phi~~.
\eeq
This choice of embedding map leads to the induced metric and worldvolume Killing vector field
\beq
\ds=-\frac{H(-1,x_R)}{H(x_R,-1)}\left(d\tau+\widehat \omega(x_R, -1)\right)^2+\frac{F(-1,x_R)}{H(-1,x_R)}d\phi^2~~,~~\ku a\partial_a=\partial_\tau+\Omega\partial_\phi~~,
\eeq
while the transverse Killing vector field is now aligned with the $\chi$ axis such that $\kpu\mu\partial_\mu=\partial_\chi$. The modulus of the surface Killing vector and the determinant of the induced metric read
\beq
\k^2=\frac{F(-1,x_R)H(x_R,-1)\Omega^2-H(-1,x_R)^2(1+\widehat\omega(x_R,-1)\Omega)^2}{H(-1,x_R)H(x_R,-1)}~~,~~\gamma=-\frac{F(-1,x_R)}{H(x_R,-1)}~~.
\eeq
The functions $H(-1,x_R)$, $H(x_R,-1)$ and $F(-1,x_R)$ are given by \eqref{eq:brfunctions} with the replacement $y_R\to x_R$ while the one form $\widehat\omega(x_R,-1)$ reads
\beq
\widehat\omega(x_R,-1)=\frac{4 R_\circ \sqrt{(\nu -1)^2} \nu  \left(x_R^2-1\right)}{\sqrt{2}\left(\sqrt{\nu}-1\right)^2 \left(\nu +2 \sqrt{\nu }-2 \nu ^{3/2} x_R^2+\nu ^2 x_R^2-\nu  x_R^2-1\right)}d\psi~~.
\eeq
With this in hand we can now look for possible extremal bi-ring configurations.

\subsubsection{Solution space}
\begin{figure}[h!] 
\centering
  \includegraphics[width=0.5\linewidth]{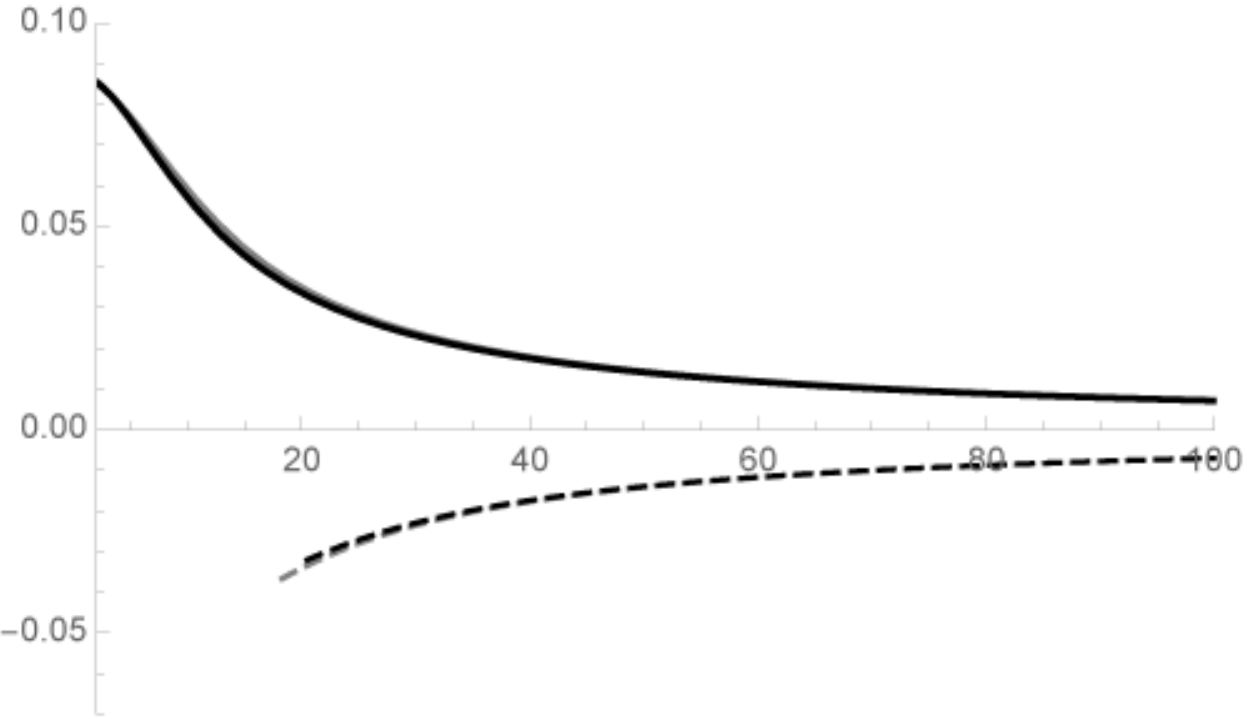}
  \begin{picture}(0,0)(0,0)
  \put(-237,115){ $\Omega $}
    \put(-20,30){ $R$}
\end{picture} 
\caption{$\Omega$ as a function of $R$ in units $R_\circ=1$ for $\nu=1/2$. The black curves describe the di-ring solution of the previous section while the grey cruves describe the bi-ring solution. The thick curves represent '+' branch while the dashed curves represent the '-' branch.}\label{fig:dibi}
\end{figure}
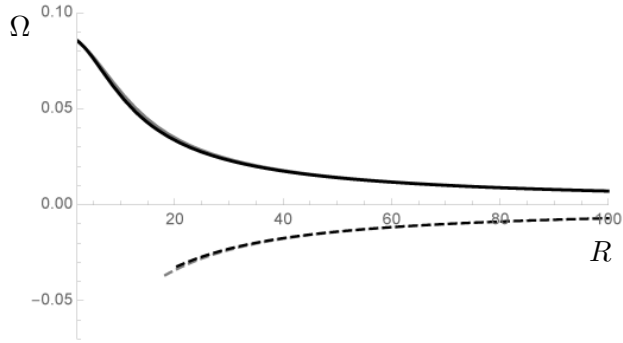

The solution space of bi-rings is qualitatively the same as the case of di-rings and it is also valid for $\widehat\Omega \ell\gg1$ with $\ell\equiv\text{min}(R,R_\circ)$. In fact, for small $\textbf{R}_\circ$ the angular velocity has a similar expansion \eqref{eq:Omegaexp} in powers of $\textbf{R}_\circ$ but with a qualitative difference: the first correction of order $\mathcal{O}\left(\textbf{R}_\circ^2\right)$ has the opposite sign. In particular, the branch of solutions that is rotating in the clockwise direction has the following expansion at small $\textbf{R}_\circ$
\beq \label{eq:Omegaexp1}
\Omega=\frac{1}{\sqrt{2}R}\left(1\pm\frac{2\sqrt{\nu}-8\nu-2\nu^{2/3}\pm\nu^2-1}{2(1-\nu)^2}\textbf{R}_\circ^2+\mathcal{O}\left(\textbf{R}_\circ^4\right)\right)~~,
\eeq
where the '+' sign denotes the di-ring solution and the '-' sign the bi-ring solution. The behaviour of the bi-ring solution for arbitrary $\textbf{R}_\circ$ is shown in Fig.~\ref{fig:dibi} and contrasted with the di-ring solution.
At large distances away from the centre black ring the solution reduces to the flat space case \eqref{eq:ring}, as expected. From Eq.~\eqref{eq:Omegaexp1} we see that bi-rings are rotating faster than di-rings for small $\textbf{R}_\circ$, since the function multiplying the factor $\textbf{R}_\circ^2$ in Eq.~\eqref{eq:Omegaexp1} is always negative for any $\nu$. From Fig.~\ref{fig:dibi} we see that this continues to be the case and becomes quantitively different for larger values of $\textbf{R}_\circ$. This is because of the fact that the centre black ring has $J_\psi<J_\chi$ for any given value of $R$. For other values of $\nu$, the solution space exhibits a similar behaviour as that presented in Fig.~\ref{fig:diring}.

\subsection{Phase diagram of extremal black holes in flat space} \label{sec:phase}
\begin{figure}[h!] 
\centering
\begin{subfigure}{.5\textwidth}
  \centering
  \includegraphics[width=1\linewidth]{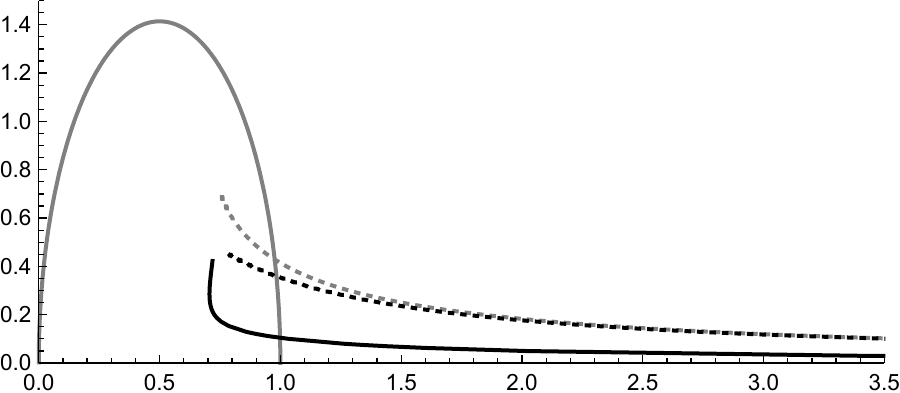}
  \begin{picture}(0,0)(0,0)
  \put(-130,95){ $a_H $}
    \put(80,7){ $j_\chi$}
\end{picture}  
\end{subfigure}%
\begin{subfigure}{.5\textwidth}
  \centering
  \includegraphics[width=0.5\linewidth]{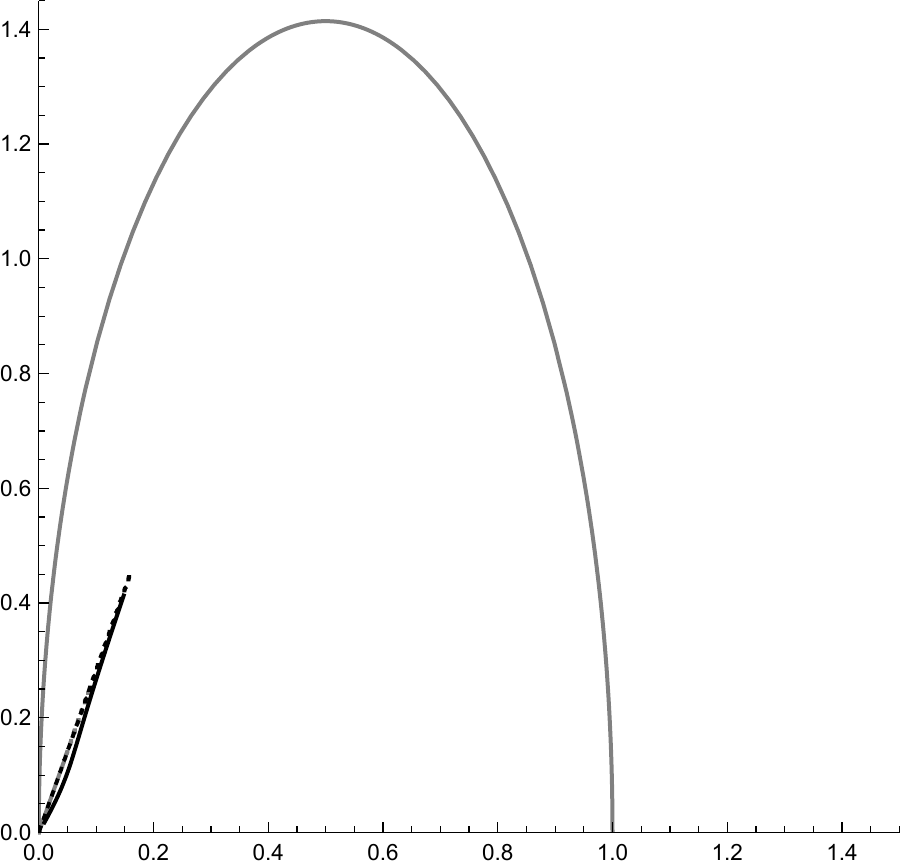}
  \begin{picture}(0,0)(0,0)
  \put(-135,70){ $a_H $}
    \put(-25,-5){ $j_\psi$}
\end{picture} 
\end{subfigure}%
\caption{The l.h.s. shows the reduced area $a_H$ as a function of the reduced angular momentum $j_\chi$ while the r.h.s. shows as a function of $j_\psi$. The grey thick line represents the exact extremal Myers-Perry black hole, the grey dotted line the exact extremal doubly-spinning back ring while the black dotted line represents the approximate doubly-spinning black ring curve using \eqref{eq:tbf} and the black line is the extremal di-ring of Sec.~\ref{sec:diring}. }  
  \label{fig:phase0}
\end{figure}
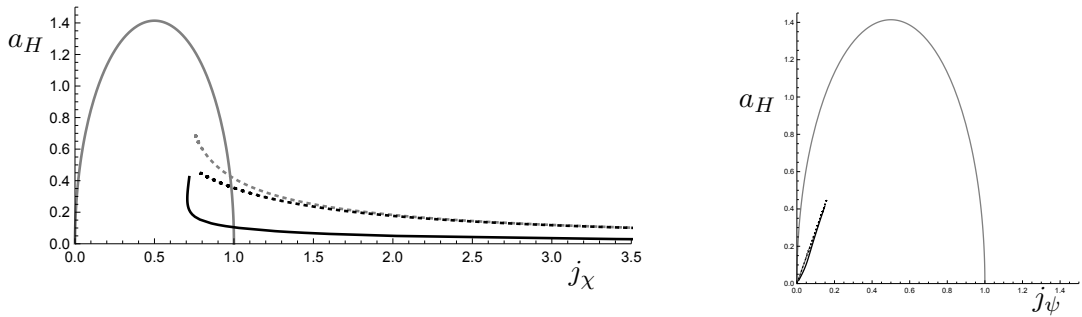

We now look at the phase diagram of extremal black holes in $D=5$ vacuum Einstein gravity. In general, the phase diagram will be quite complex but it is possible to consider the simplest case, analogous to \cite{Emparan:2010sx}, where the extremal solutions are in thermodynamic equilibrium. This implies that the angular velocity of each horizon must be equal, i.e.
\beq \label{eq:thermoeq}
\Omega_\chi=\Omega~~,~~\Omega_\psi=\widehat\Omega~~,
\eeq
since the temperature at each horizon vanishes. Using the thermodynamic formulae of Sec.~\ref{sec:thermo}, taking into account the additional binding energy of the outermost horizon, we can solve numerically for the condition \eqref{eq:thermoeq}. The phase diagram is presented in the dimensionless variables introduced in \cite{Emparan:2007wm}
\beq
a_H=\frac{3}{16}\sqrt{\frac{3}{\pi}}\frac{A_H^T}{(G M_T)^{3/2}}~~,~~j_\chi=\sqrt{\frac{27\pi}{32}}\frac{J^T_\chi}{(GM_T)^{3/2}}~~,~~j_\psi=\sqrt{\frac{27\pi}{32}}\frac{J^T_\psi}{(GM_T)^{3/2}}~~.
\eeq
It can be explicitly verified that, in the approximation scheme considered throughout this paper, the black saturn solution of Sec.~\ref{sec:bs} and the bi-ring solution of Sec.~\ref{sec:biring} do not exhibit solutions with \eqref{eq:thermoeq}. In the black saturn case, note that we need to require $\widehat \Omega m\gg1$ and since we have identified $\widehat\Omega=1/m$ by virtue of Eq.~\eqref{eq:angMP}, this requirement cannot be satisfied. In addition, the equivalence $\widehat\Omega=1/m$ only holds when the ring horizon coincides with the Myers-Perry horizon for which the solution breaks down. In the bi-ring case, solutions can be found but the requirement $\widehat\Omega R\gg1$ cannot be satisfied given that we have identified $\widehat\Omega=\Omega_\chi$ where $\Omega_\chi$ is given by \eqref{eq:omegabr}. The di-ring  solution of Sec.~\ref{sec:diring}, however, does admit valid solutions within this approximation, as shown in Fig.~\ref{fig:phase0}.

In Fig.~\ref{fig:phase0} we display the behaviour of phase space in thermodynamic equilibrium in both rotation planes for the exact extremal Myers-Perry black hole, the exact extremal doubly-spinning black ring  as well as the approximate extremal doubly-spinning black rings of Sec.~\ref{sec:extremaldoubly} and the extremal di-rings of Sec.~\ref{sec:diring}. From Fig.~\ref{fig:phase0} we see that the curve of the approximate extremal doubly-spinning black ring is in good agreement with the exact curve up to values of $\widehat\Omega R\sim 0.1$ in the $(a_H,j_\chi)$ diagram. In the $(a_H,j_\psi)$ diagram, the curves are almost superimposed for all values of $\widehat\Omega R$. This is because, as noted in Sec.~\ref{sec:extremaldoubly}, the mass and angular momentum $j_\psi$ do not receive corrections in $\widehat\Omega R$. The 
rest of the phase space has the expected behaviour, where the di-ring solution has lower entropy in the $(j_\chi,a_H)$ diagram, analogous to the finite temperature case \cite{Emparan:2010sx}. It should be noted that, as mentioned in Sec.~\ref{sec:diring}, the approximation is valid if $\widehat \Omega \ell\gg1$ with $\ell=\text{min}\left(R,R_\circ\right)$. Since we have identified $\widehat\Omega$ with $\Omega_\psi$ in \eqref{eq:omegabr}, we must require
\beq
\frac{1}{2\sqrt{2}}\left(\frac{(1-\sqrt{\nu})(1+\nu)}{\nu+\sqrt{\nu}}\right)\gg1~~.
\eeq 
This implies that the particular case of thermal equilibrium is only valid for $0<\nu\lesssim1/1000$. In turn, this implies that we should only trust the behaviour of $j_\chi$ for di-rings as presented in Fig.~\ref{fig:phase0} for $j_\chi \gtrsim 1.2$ and similarly only for $j_\psi\lesssim 4/100$.

The condition for thermodynamic equilibrium \eqref{eq:thermoeq} eliminates the two-parameter degeneracy of the solutions constructed here and this is why, once this condition is implemented, the phase diagram for the di-ring solution, as shown in Fig.~\ref{fig:phase0}, consists of a single curve. In general, the solutions constructed here trace out a plane in phase space. 

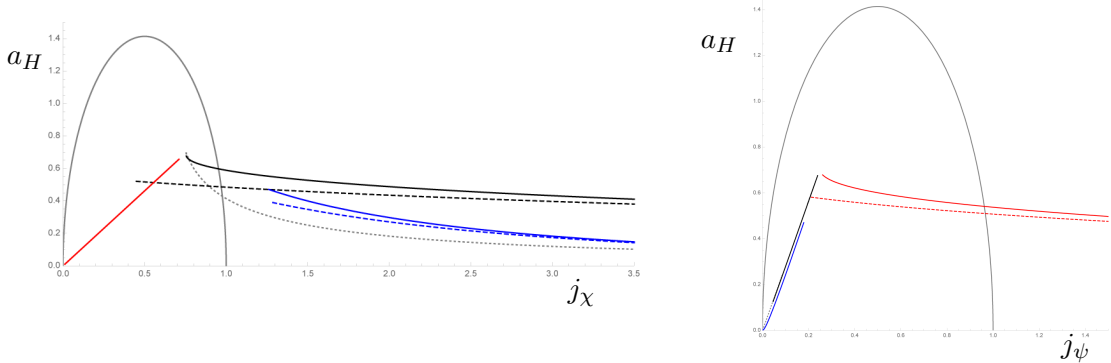
\begin{figure}[h!] 
\centering
\begin{subfigure}{.5\textwidth}
  \centering
  \includegraphics[width=1\linewidth]{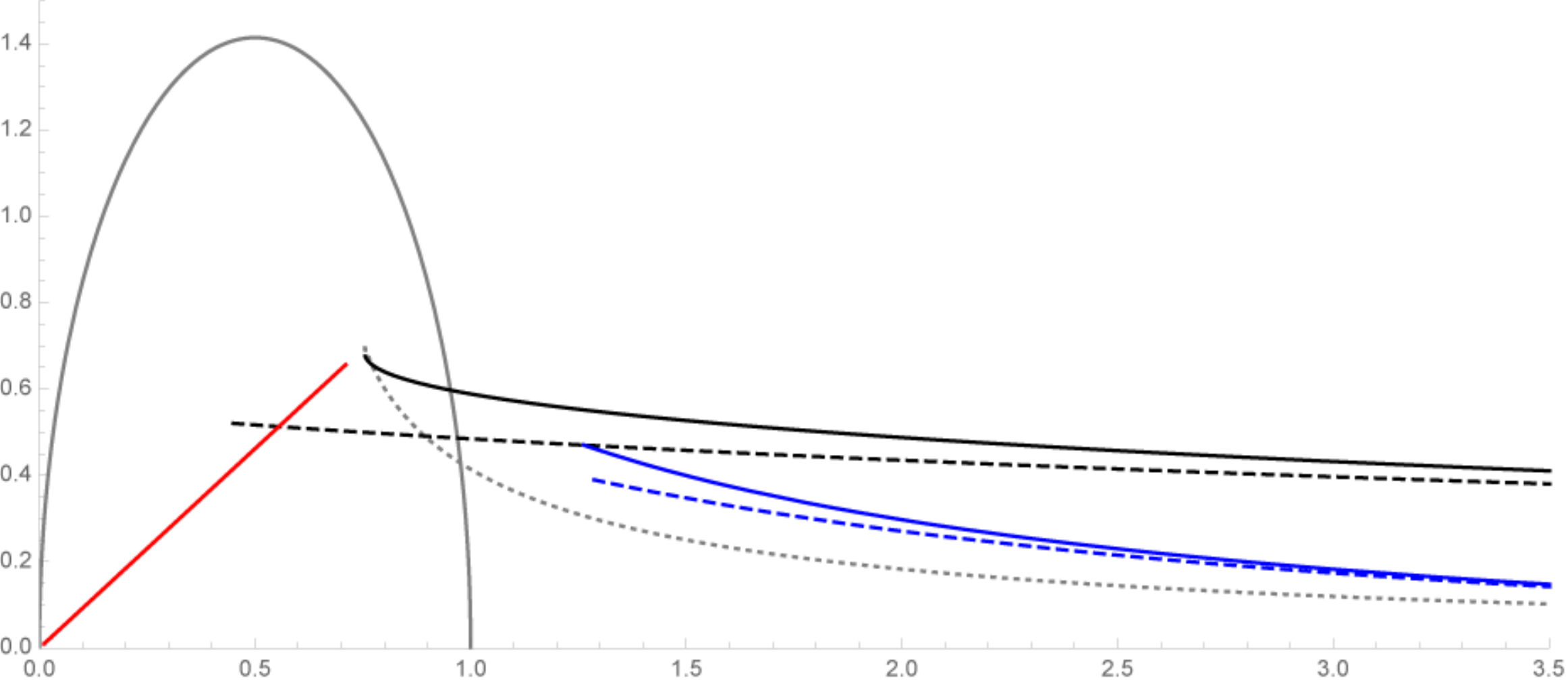}
  \begin{picture}(0,0)(0,0)
  \put(-130,95){ $a_H $}
    \put(80,7){ $j_\chi$}
\end{picture}  
\end{subfigure}%
\begin{subfigure}{.5\textwidth}
  \centering
  \includegraphics[width=0.6\linewidth]{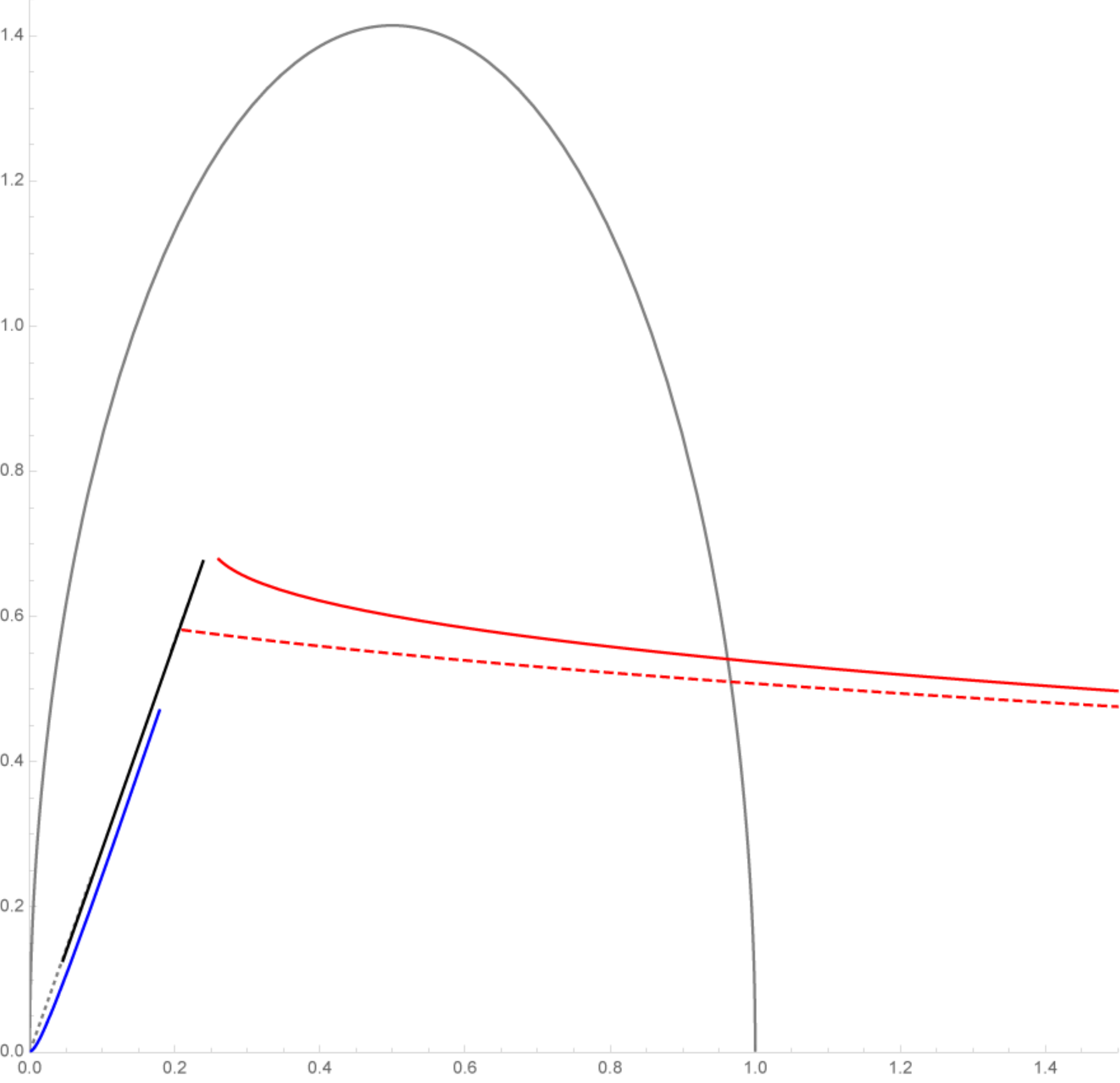}
  \begin{picture}(0,0)(0,0)
  \put(-160,110){ $a_H $}
    \put(-25,-5){ $j_\psi$}
\end{picture} 
\end{subfigure}%
\caption{On the l.h.s. it is shown the reduced area $a_H$ as a function of the reduced angular momentum $j_\chi$ while on the r.h.s. it is shown as a function of $j_\psi$. The grey line represents the exact extremal Myers-Perry black hole while the grey dotted line the exact extremal doubly-spinning back ring. The remaining thick curves represent the '+' branch of the different solutions while the dashed curves represent the '-' branches. The black lines represent the di-ring with $R_\circ=1$, $\widehat\Omega=10$ and $\nu=1/2$ and the red lines the bi-ring solution for the same values of the parameters. The blue lines represent the black saturn solution for equal rotation parameters and $m=1$, $\widehat\Omega=10$ with $a=1/2$.}  
  \label{fig:phase1}
\end{figure}

In Fig.~\ref{fig:phase1} we show different curves, for given values of $\nu$ or rotation parameters $a,b$, for both branches of solutions. The solutions constructed above require $\left(\widehat\Omega R_\circ,\widehat\Omega m\right)\gg1$ as well as $\widehat \Omega \ell\gg1$ for $\ell=\text{min}\left(R_\circ,R\right)$ or $\ell=\text{min}\left(m,R\right)$. In the phase diagram of Fig.~\ref{fig:phase1} we have kept $\widehat\Omega R_\circ=\widehat\Omega m=10$ and $\widehat\Omega R\gtrsim 100$ as not to extend the curves beyond their regime of validity.

The outermost horizon of all the novel solutions with disconnected horizons presented here is ultraspinning in the $\chi$ direction and has finite rotation in the transverse $\psi$ direction, except the bi-ring solution which has finite rotation in the $\chi$ direction and is ultraspinning in the $\psi$ direction. It is therefore expected that the angular momentum of most of these solutions is unbounded in the $\chi$ direction and bounded in the $\psi$ direction due to the extremal bound of the Kerr string. Indeed this is the case, as can be seen in both plots in Fig.~\ref{fig:phase1}. On the other hand, the angular momentum of the bi-ring solution is unbounded in the $\psi$ direction and bounded in the $\chi$ direction. The boundedness of the bi-ring solution in the $\chi$ direction is not only explained by the extremal bound on the Kerr string but also by the fact that the mass of the outer ring grows relatively fast with the angular momentum compared to the angular momentum of the inner ring horizon. The bi-ring solution includes cases in which the inner ring is ultraspinning in the $\chi$ direction, even though the outer ring is not. If we would define a dimensionless quantity $j_\chi$ associated with the inner horizon, $j_\chi$ would be unbounded. However, taking into account the mass and angular momentum of the outer ring, the total angular momentum is not bounded in the $\chi$ direction. It is also clear from Fig.~\ref{fig:phase1} that the di-ring and the bi-ring solution show qualitatively the same features but in the opposite rotation planes.


\section{Extremal black holes in $D\ge6$} \label{sec:bhd}
In this section we look at certain classes of extremal black holes in $D\ge6$. As mentioned in Sec.~\ref{sec:kerrbrane}, the extremal version of the isolated horizons found in $D\ge6$ using non-spinning branes \cite{Emparan:2009vd, Armas:2015kra, Armas:2015nea} are regular solutions to \eqref{eq:eom1}, provided there are no symmetry restrictions. This implies that, for instance, extremal versions of the black cylinders and black odd-spheres of \cite{Emparan:2009vd} yield regular horizons. In this section we analyse the case of a configuration representing a single horizon, namely that corresponding to the extremal Myers-Perry black hole for which comparison with exact analytic results can be made. We will show that the extremal Myers-Perry black hole with one finite angular momentum and several ultraspinning ones can be captured by the effective theory introduced here. In addition, we provide a generalisation of the black saturn solution of Sec.~\ref{sec:bs} in which an horizon with topology $\mathbb{S}^{3}\times\mathbb{S}^{2}$ is wrapped around the $\mathbb{S}^{5}$ horizon of the Myers-Perry black hole in $D=7$.



\subsection{Extremal Myers-Perry black holes} \label{sec:MPX}
In this section we construct extremal Myers-Perry black holes with one finite angular momenta and compare it against the exact solution finding complete agreement. We begin with the $D=6$ case where the black hole only carries two angular momenta. We then move on to the general case of arbitrary even dimension.

\subsubsection{Extremal Myers-Perry black hole in $D=6$}
Consider Minkowski space in $D=6$ written in the form
\beq
ds^2=-dt^2+dr_1^2+r_1^2d\chi^2+dr_2^2+r_2^2d\psi^2+dx^2~~,
\eeq 
where $0\le r_1,r_2<\infty$ and $0\le \chi,\psi\le2\pi$. We now embed a disc geometry by choosing the embedding map
\beq
t=\tau~~,~~r_1=\rho~~,~~\chi=\phi~~,~~r_2=0~~,~~x=0~~.
\eeq
Setting the disc to rotate with angular velocity $\Omega$ along the $\chi$ direction and with transverse angular velocity $\widehat\Omega$ along the $\psi$ direction, the induced metric and Killing vector fields become
\beq
\ds=-d\tau^2+d\rho^2+\rho^2d\phi^2~~,~~\ku a\partial_a=\partial\tau+\Omega\partial_\phi~~,~~\textbf{k}^{\mu}_\perp\partial_\mu=\widehat\Omega\partial_\psi~~.
\eeq
This embedding is a trivial solution of \eqref{eq:eom} since the extrinsic curvature vanishes and the background space is flat. The non-trivial feature of this embedding is the existence of its boundary located at $\rho=\rho_+$ for which $\k^2(\rho_+)=1-\Omega^2\rho_+^2=0$ such that
\beq
\rho_+=\frac{1}{\Omega}~~.
\eeq
As mentioned in Sec.~\ref{sec:regval}, configurations with boundaries deserve special attention, as to describe the physics accurately at the boundary one must employ a boundary expansion due to \eqref{eq:edge}. In order to explicitly see this, we consider the length scales associated with higher order corrections \eqref{eq:val}. The only nontrivial scales are those associated with the acceleration and vorticity of the fluid.  We evaluate
\beq
|\mathfrak{a}^{b}\mathfrak{a}_{b}|^{-\frac{1}{2}}=\frac{\k^2}{\rho\Omega^2}~~,~~|\omega_{ab}\omega^{ab}|^{-\frac{1}{2}}=\frac{\k^2}{\sqrt{2}\Omega}~~.
\eeq
Both the scales imply that near the boundary for which $\k=0$, the requirement \eqref{eq:val} cannot be satisfied and the approximation breaks down, while near the axis of rotation ($\rho=0$), the scale associated with vorticity implies that $\Omega/\widehat\Omega\ll1$. In general, one must require
\beq \label{eq:req}
\frac{\Omega}{\widehat\Omega}\ll\textbf{k}~~,
\eeq
over the disc. This implies that the disc must be rotating faster in the transverse direction near the boundary in order to satisfy \eqref{eq:req}. 

Keeping this in mind, we introduce the cut-off $\epsilon$ and integrate the free energy \eqref{eq:eff0} in the interval $0\le\rho\le(\rho_+-\epsilon)$ for small $\epsilon$. We find the on-shell free energy
\beq
\mathcal{F}[\Omega,\widehat\Omega]|_{T=0}=\frac{\pi}{6G\widehat\Omega\Omega^2}+\mathcal{O}\left(\epsilon^{\frac{3}{2}}\right)~~.
\eeq
Even though we have introduced the cut-off $\epsilon$, the on-shell free energy is independent of it at leading order in a boundary expansion. This is because the majority of the mass is concentrated in the centre of the disc. Using the free energy, together with the formulae of Sec.~\ref{sec:thermo}, we can obtain the remaining thermodynamic properties, including the mass, angular momenta and entropy
\beq \label{eq:Mpmp}
M=\frac{2\pi}{3G\widehat\Omega\Omega^2}~~,~~J_\chi=\frac{\pi}{3G\widehat\Omega\Omega^3}~~,~~J_\psi=\frac{\pi}{6G\widehat\Omega^2\Omega^2}~~,~~S=\frac{\pi^2}{3G\widehat\Omega^2\Omega^2}~~,~~\boldsymbol{\mathcal{T}}=0~~. 
\eeq
Before comparing these results with the exact solution, we will generalise them to any even $D$.

\subsubsection{Extremal Myers-Perry black hole in even $D\ge6$}
We now generalise the above results to include Myers-Perry black holes with an arbitrary number $l$ of ultraspins in even $D$ dimensions. We write $D$-dimensional Minkowski space as
\beq
ds^2=-dt^2+\sum_{i=1}^{l}\left(dr_i^2+r_i^2d\chi_i^2\right)+dr_{l+1}^2+r_{l+1}^2d\psi^2+dx^2~~,
\eeq
with $D=2l+4$. The fact that $D$ is always even supervenes on the fact that the effective theory employed here is only valid for $n=1$. In this background we embed a $2l+1$ ellipsoid rotating along the $l$ angular directions $\chi_i$ and transverse to it in the $\psi$ direction by choosing the embedding map
\beq
t=\tau~~,~~r_i=\rho_i~~,~~\chi_i=\phi_i~~,~~r_{l+1}=0~~,~~x=0~~,
\eeq
giving rise to the induced metric and Killing vector fields
\beq
\ds=-d\tau^2+\sum_{l}\left(d\rho_i^2+\rho_i^2d\phi_i^2\right)~~,~~\ku a\partial_a=\partial_\tau+\sum_{l}\Omega^{(l)}\partial_{\phi_{(l)}}~~,~~\textbf{k}^{\mu}_\perp\partial_\mu=\widehat\Omega\partial_\psi~~.
\eeq
The boundary of the ellipsoid is now located at $\sum_{i=1}^l\Omega_{(l)}^2\rho_{(l)}^2=0$, where, by evaluating the scales \eqref{eq:val}, the approximation is expected to break down. In general, \eqref{eq:val} requires that
\beq \label{eq:valMP}
\frac{\sqrt{\left(\sum_{i=1}^l \Omega_{(l)}^2\right)}}{\widehat\Omega}\ll \textbf{k}~~.
\eeq
As in the $D=6$ case, introducing a cut-off $\epsilon$ does not change the thermodynamics at zeroth order in a boundary expansion. We thus compute the on-shell free energy
\beq \label{eq:fmpbf}
\mathcal{F}[\Omega_{(l)},\widehat\Omega]|_{T=0}=\frac{1}{8\widehat\Omega G}\frac{\pi^{(l+\frac{1}{2})}}{\Gamma\left(l+\frac{3}{2}\right)}\prod_{i=1}^{l}\Omega_{(l)}^{-2}+\mathcal{O}\left(\epsilon^{\frac{3}{2}}\right)~~.
\eeq
From here we determine the remaining thermodynamic properties
\beq
M=\frac{(l+1)}{4\widehat\Omega G}\frac{\pi^{(l+\frac{1}{2})}}{\Gamma\left(l+\frac{3}{2}\right)}\prod_{i=1}^{l}\Omega_{(l)}^{-2}~~,~~J_{i}=\frac{1}{4\widehat\Omega \Omega_{i} G}\frac{\pi^{(l+\frac{1}{2})}}{\Gamma\left(l+\frac{3}{2}\right)}\prod_{i=1}^{l}\Omega_{(l)}^{-2}~~,
\eeq
\beq \label{eq:smpbf}
J_{i}=\frac{1}{4\widehat\Omega \Omega_{i} G}\frac{\pi^{(l+\frac{1}{2})}}{\Gamma\left(l+\frac{3}{2}\right)}\prod_{i=1}^{l}\Omega_{(l)}^{-2}~~,~~S=\frac{\pi}{4\widehat\Omega^2 G}\frac{\pi^{(l+\frac{1}{2})}}{\Gamma\left(l+\frac{3}{2}\right)}\prod_{i=1}^{l}\Omega_{(l)}^{-2}~~,~~\boldsymbol{\mathcal{T}}=0~~.
\eeq


\subsubsection{Comparison with the extremal Myers-Perry solution}
In order to compare the thermodynamics above with the exact extremal Myers-Perry solution in even $D$, we follow the conventions of \cite{Gibbons:2004ai}.\footnote{We have exchanged $m$ appearing in \cite{Gibbons:2004ai} with $\mu$ appearing here.} The Myers-Perry black hole in even $D$ dimensions is described by $N$ planes of rotation such that $D=2N+2$. We single out one plane of rotation by redefining $N$ such that $N=l+1$. Each plane of rotation has an associated rotation parameter $a_i=1,..,l+1$. We take the first $l$ parameters to be ultraspinning such that $a_i\gg r_+~,~i=1,..,l$ where $r_+$ is the outer horizon radius and assumed to be finite \cite{Emparan:2003sy}. The last rotation parameter $a_{l+1}$ remains finite. In this case, there exists an inner and outer horizon $r_\pm$ such that
\beq
r_\pm=\frac{\mu\pm\sqrt{\mu^2-a_{l+1}^2(\prod_{i=1}^{l} a_i^2)^2}}{\prod_{i=1}^{l} a_i^2}~~.
\eeq 
The Myers-Perry black hole is extremal if the two horizons coincide, this implies that
\beq
\mu=a_{l+1}\prod_{i=1}^{l} a_i^2~~,~~r_+=a_{l+1}~~,
\eeq
and hence the angular velocities read
\beq
\Omega_{i}=\frac{1}{a_i}~,~i=1,...,l ~~,~~\Omega_{l+1}=\frac{1}{2a_{l+1}}~~.
\eeq
In order to compare it with the thermodynamic properties obtained above it is sufficient to verify that the free energy and the entropy match, since from the free energy the remaining properties can be obtained. 
Given the limit above, the free energy and entropy in \cite{Gibbons:2004ai} read\footnote{Here $\mathcal{F}|_{T=0}=T I_D$ where $I_D$ is the Euclidean action in \cite{Gibbons:2004ai}. }
\beq
\mathcal{F}|_{T=0}=\frac{1}{8\Omega_{l+1} G}\frac{\pi^{(l+\frac{1}{2})}}{\Gamma\left(l+\frac{3}{2}\right)}\prod_{i=1}^{l}\Omega_{i}^{-2}~~,~~S|_{T=0}=\frac{\pi}{4\Omega^2_{l+1} G}\frac{\pi^{(l+\frac{1}{2})}}{\Gamma\left(l+\frac{3}{2}\right)}\prod_{i=1}^{l}\Omega_{i}^{-2}~~.
\eeq
Therefore, we have perfect agreement with \eqref{eq:fmpbf} and \eqref{eq:smpbf} provided we identify $\Omega_i=\Omega_{(l)}~,~i=1,...,l$ and $\Omega_{l+1}=\widehat\Omega$.
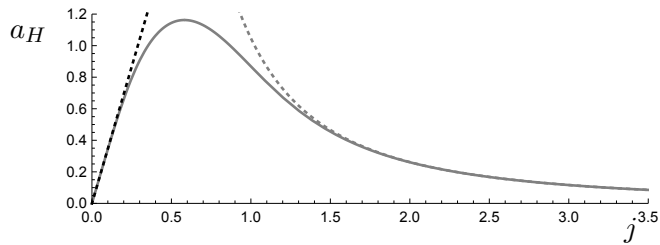
\begin{figure}[h!] 
\centering
  \includegraphics[width=0.5\linewidth]{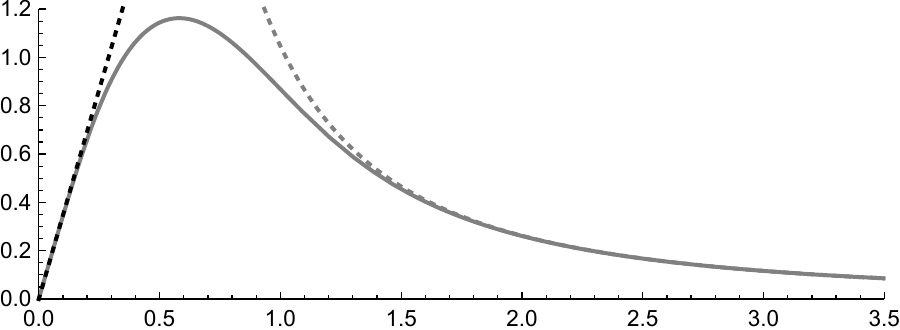}
  \begin{picture}(0,0)(0,0)
  \put(-250,70){ $a_H $}
    \put(-20,-5){ $j$}
\end{picture} 
\caption{The reduced $a_H$ as a function of the reduced angular momentum $j$. The thick grey line represents the exact extremal Myers-Perry black hole in $D=6$, which exhibits the same behaviour, due to spherical symmetry, in both rotation planes. The dashed grey line is the angular momentum $j_\chi$ of the approximate extremal Myers-Perry \eqref{eq:Mpmp} while the black dashed line is the angular momentum $j_\psi$.} \label{fig:MPD6}
\end{figure}
 One may compare the approximate solution in $D=6$ given by \eqref{eq:Mpmp} with the exact extremal Myers-Perry solution. Using the thermodynamic properties of the Myers-Perry solution given in \cite{Gibbons:2004ai} and reduced quantities analogous to \eqref{eq:thermoeq} as in \cite{Emparan:2007wm}, the phase diagram can be depicted as in Fig.~\ref{fig:MPD6}. We see that the Myers-Perry solution obtained using blackfold methods approximates well the exact solution for $j_\chi\gtrsim1.5$ and for $j_\psi\lesssim0.5$. It is clear that the blackfold approximation works better than expected given that the regime of validity \eqref{eq:valMP} indicates that one should only expect good agreement if $j_\chi\gg1$. This gives further evidence for the correctness of the effective theory employed here.


\subsection{Extremal higher-dimensional black saturns} \label{sec:highblacks}
In this section we construct a generalisation of the black saturn solution of Sec.~\ref{sec:bs} for which an horizon with $\left(\mathbb{S}^{3}\times \mathbb{S}^{2}\right)$ is placed around a centre black hole and we briefly analyse its solution space. Further investigation, including their finite temperature versions, will be carried out in \cite{toappear}. The complete solution has horizon topology $\mathbb{S}^{5}\cup\left(\mathbb{S}^{3}\times \mathbb{S}^{2}\right)$. In order to do so, we take a Myers-Perry black hole in $D=7$ as the background and for simplicity we focus on the case in which all three rotation parameters are equal to each other $a_i=a~,~i=1,2,3$. The metric of the Myers-Perry black hole can be written as
\beq \label{eq:dsmp7}
ds^2=-dt^2+\sum_{i=1}^{3}\left(r^2+a^2\right)\left(d\mu_i^2+\mu_i^2d\chi_i^2\right)+\frac{m^4r^2}{\Pi F}\left(dt-a\sum_{i=1}^{3}\mu_i^2d\chi_i\right)^2+\frac{\Pi F}{\Pi-r^2m^4}dr^2~~,
\eeq
where we have introduced the functions
\beq
F(r,\mu_i)=1-\frac{a^2}{r^2+a^2}~~,~~\Pi(r)=\left(r^2+a^2\right)^3~~.
\eeq
The coordinate $r$ runs lies in the interval $r_+<r<\infty$, where $r_+$ is the location of the horizon. The coordinates $\chi_i$ are azimuthal angles $0\le\chi_i\le2\pi$ while the direction cosines lie in the interval $-1\le\mu_i\le1$ and satisfy the constraint $\mu_1^2+\mu_2^2+\mu_3^2=1$.

The horizon $r_+$ is given by the outermost real root of $\Pi(r_+)-m^4r_+^2=0$. At extremality, this implies that
\beq
r_+=\frac{a}{\sqrt{2}}~~,~~m=\frac{3^{3/4}}{\sqrt{2}}a~~,
\eeq
in which case all the angular velocities are rotating clockwise with magnitude
\beq
\Omega_{\chi_i}=\frac{\sqrt{2}}{3^{1/4}m}~~.
\eeq
We will now place a $3+1$-dimensional submanifold with transverse spin rotating in this background.


\subsubsection{Embedding and solution space}
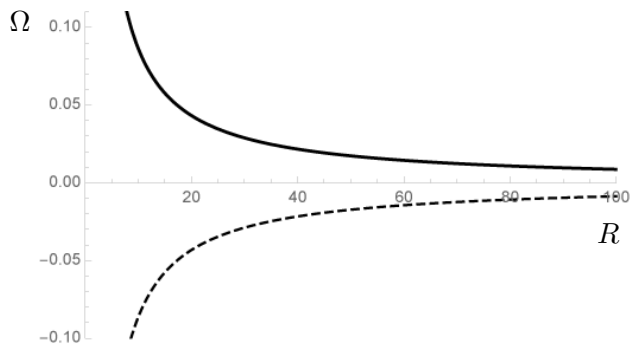
\begin{figure}[h!] 
\centering
  \includegraphics[width=0.5\linewidth]{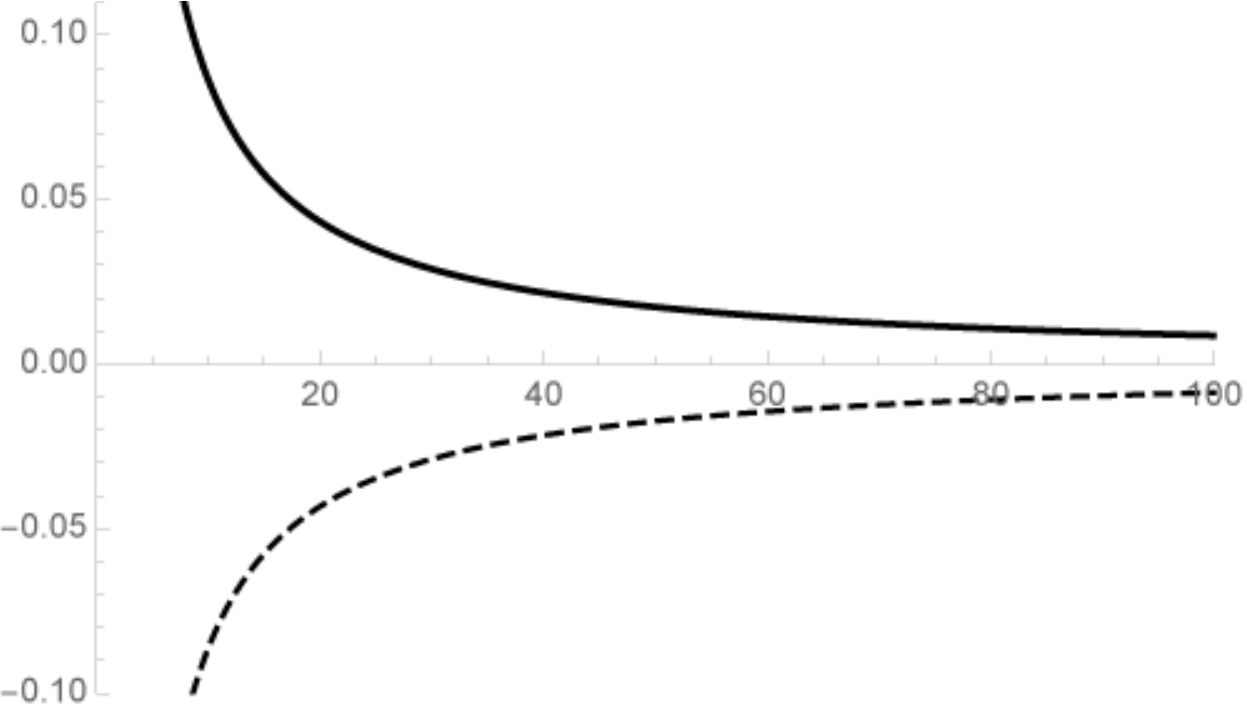}
  \begin{picture}(0,0)(0,0)
  \put(-240,120){ $\Omega $}
    \put(-20,40){ $R$}
\end{picture} 
\caption{$\Omega$ as a function of $R$ in units $m=1$. The black line is the '+' branch while the dashed line is the '-' branch.} \label{fig:BSequalD}
\end{figure}

We embedd a $4$-dimensional submanifold in the background \eqref{eq:dsmp7} by choosing the embedding map
\beq
t=\tau~~,~~\mu_1=\sin\theta~~,~~\mu_3=0~~,~~\chi_1=\phi_1~~,~~\chi_2=\phi_2~~,~~r=R~~,
\eeq
for which $0\le\theta\le\pi$, and we set it to rotate with equal angular velocity $\Omega$ along the $\chi_1,\chi_2$ directions and will transverse angular velocity $\widehat\Omega$ in the $\chi_3$ direction. The induced metric reads
\beq
\begin{split}
\ds=-d\tau^2&+(R^2+a^2)\left(d\theta^2+\cos^2\theta d\phi_1^2+\sin^2\theta d\phi_2^2\right)\\
&+\frac{m^4R^2}{\Pi(R) F(R,\theta)}\left(d\tau-a\cos^2\theta d\phi_i-a\sin^2\theta d\phi_2\right)^2~~,
\end{split}
\eeq
while the Killing vector fields take the form
\beq
\ku a\partial_a=\partial_\tau+\Omega\left(\partial_{\phi_1}+\partial_{\phi_2}\right)~~,~~\textbf{k}^{\mu}_\perp\partial_\mu=\widehat\Omega \partial_{\chi_3}~~.
\eeq
By evaluating $\gamma,\k$, plugging it into the free energy \eqref{eq:eff0} and varying it with respect to $R$, one obtains the equilibrium condition for $\Omega$. It can be solved exactly and it exhibits a similar behaviour as in the case of black saturns in $D=5$ with equal rotation parameters of Sec.~\ref{sec:bsequal}. It is composed of two branches, for clockwise and anti-clockwise rotation. For small $\m=m/R$, the solution exhibits the behaviour
\beq
\Omega=\Omega_0\left(1-\frac{\m^2}{3\sqrt{3}}+\mathcal{O}\left(\m^4\right)\right)~~,~~\Omega_0=\pm\frac{\sqrt{3}}{2R}~~.
\eeq
Here the zeroth order result $\Omega_0$ agrees with that of a $(n,p)=(1,3)$ odd-sphere in Minkowski space as found in \cite{Emparan:2009vd}. In Fig.~\ref{fig:BSequalD} we exhibit the complete behaviour of $\Omega$ as a function of $R$. 
The behaviour of $\Omega$ is very similar to that presented in Fig.~\ref{fig:BSequal}. The '+' branch terminates at $R=r_+$ while the '-' branch terminates at a slightly higher value of $R$. The approximation is valid when $\widehat\Omega \ell \gg 1$ for $\ell=\text{min}\left(R,m\right)$ and $\widehat\Omega m\gg1$ for lower values of $R$.

%

\section{Concluding remarks} \label{sec:conclusions}
Using the blackfold effective theory, we have shown the existence of new extremal solutions with disconnected horizons and reproduced existing ones. In particular, in $D=5$ these include extremal black saturns, di-rings and bi-rings, where each horizon that composes these solutions has two independent angular momenta.

All the extremal solutions found here in $D\ge6$ are non-static, agreeing with the result of \cite{Khuri:2017fun} in vacuum Einstein gravity. In addition, the topology of all horizons in $D=5$ encountered in this work is either $\mathbb{S}^{3}$ or $\mathbb{S}^{1}\times\mathbb{S}^{2}$, thus agreeing with the extremal near-horizon geometry classification of \cite{Kunduri:2008rs, Kunduri:2013ana} and the uniqueness theorems of \cite{Hollands:2007aj, Figueras:2009ci}. It would be interesting to obtain extremal versions of the black holes found in \cite{Armas:2010hz, Armas:2015kra, Armas:2015qsv} in (Anti)-de Sitter spacetime and to check whether they also agree with \cite{Khuri:2017zqg, Khuri:2017fun}. 

In Sec.~\ref{sec:phase}, the phase diagram of $D=5$ extremal black holes was analysed. It was shown that imposing thermodynamic equilibrium as in \cite{Emparan:2010sx} leads to the exclusion of the black saturn and bi-ring solution. The fact that there are no solutions in thermal equilibrium for the extremal black saturn and bi-rings does not necessarily mean that such phases of thermal equilibrium do not exist for the complete solution. However, it does mean that they will not have blackfold limits. This is somewhat surprising and it hints at the fact that extremal disconnected horizons are quite constrained systems. Their finite temperature counterparts are expected to exhibit such phases \cite{toappear} and indeed they do at least when only rotating in one of the rotation planes \cite{Emparan:2010sx}. This will be further investigated in a future publication \cite{toappear}.

The application of the effective theory presented here was restricted to solving the set of constraint equations, which are a necessary condition for the existence of black hole solutions. We have not constructed any explicit metrics. Showing that solving this set of constraint equations is a necessary condition supervenes on a simple generalisation of the arguments of \cite{Camps:2012hw} and will be spelled out in detail in \cite{toappear}. Constructing the explicit metrics requires the more evolved task of generalising \cite{Emparan:2007wm} to Kerr branes. Nevertheless, we look forward to completing such task.

The solutions constructed here were restricted to $n=1$ which implied that only one of black hole angular momenta is finite. To generalise these results to $n>1$, it is necessary to work with an effective theory of Myers-Perry branes with an arbitrary number of rotation parameters. This effective theory is that of a multi-charged fluid. In a forthcoming publication \cite{toappear}, we will introduce these effective theories and study some of their stability properties including at extremality. Many new black holes with finite temperature will be constructed and, in some cases, corrections to the effective theory, following \cite{Armas:2013hsa, Armas:2013goa, Armas:2014bia, Armas:2014rva}, will be considered. 

Making use of the classification of extremal near-horizon geometries, Ref.~\cite{Figueras:2008qh} conjectured the exact thermodynamic properties for a generalisation of the black ring solution of Sec.~\ref{sec:extremaldoubly}, namely, one for which the horizon topology is $\mathbb{S}^{1}\times \mathbb{S}^{D-3}$. Using the blackfold effective theory applied to Kerr branes, these solutions are not possible to construct. Instead, only solutions analogous to that surrounding the $D=7$ Myers-Perry black hole in Sec.~\ref{sec:highblacks} can be constructed. These solutions, referred to as black odd-spheres in \cite{Emparan:2009vd}, have topology $\mathbb{S}^{D-4}\times\mathbb{S}^{2}$. In order to construct the solutions of \cite{Figueras:2008qh}, one must use the effective theory of Myers-Perry branes with several non-zero transverse spins.

An interesting generalisation of the work presented here is to consider Kerr-Newmann branes, which, from the point of view of the effective theory \eqref{eq:eff}, will carry transverse spin as well as a magnetic moment in accordance with \cite{Armas:2012ac, Armas:2013aka}. Such effective theory allows to find novel extremal black hole solutions with electric charge in Einstein-Maxwell theory, including doubly-spinning charged black rings in $D=5$.\footnote{In supergravity many new black hole solutions were found using the blackfold approach \cite{Emparan:2011hg, Caldarelli:2010xz, Grignani:2010xm, Armas:2016mes}.} This direction will be pursued in a future publication.

\section*{Acknowledgements}
We would like to thank James Lucietti for useful e-mail correspondence and for bringing Ref.~\cite{Figueras:2008qh} to our attention. We would also like to thank an anonymous referee for valuable comments to this manuscript. The work of JA is supported by the ERC Starting Grant 335146 HoloBHC. TH and NO acknowledge support from the Independent Research Fund Denmark grant number DFF-6108-00340 ''Towards a deeper understanding of black holes with non-relativistic holography".

\addcontentsline{toc}{section}{References}
\footnotesize
\providecommand{\href}[2]{#2}\begingroup\raggedright\endgroup

\end{document}